\documentclass{aa}

\usepackage[varg]{txfonts}
\usepackage{adjustbox,lipsum}

\usepackage{graphicx}
\usepackage{amsmath}
\usepackage[normalem]{ulem}
\usepackage[config, labelfont={rm,bf}, font=footnotesize, singlelinecheck=false]{caption, subfig}

\usepackage{multirow}

\begin{document}

\title{The intrinsic shape of bulges in the CALIFA survey}

\author{L.~Costantin\inst{1}\thanks{luca.costantin@studenti.unipd.it}
\and J.~M\'endez-Abreu\inst{2, 3}
\and E.~M.~Corsini\inst{1, 4}
\and M.~C.~Eliche-Moral\inst{2}
\and T.~Tapia\inst{5}
\and L.~Morelli\inst{1, 4}
\and \\E.~Dalla~Bont\`a\inst{1, 4}
\and A.~Pizzella\inst{1, 4}
}

\institute {Dipartimento di Fisica e Astronomia `G. Galilei', Universit\`a di Padova, vicolo dell'Osservatorio 3, I-35122 Padova, Italy
\and Instituto de Astrof\'isica de Canarias, Calle V\'ia L\'actea s/n, E-38200 La Laguna, Tenerife, Spain
\and Departamento de Astrof\'isica, Universidad de La Laguna, Calle Astrof\'isico Francisco S\'anchez s/n, E-38205 La Laguna, Tenerife, Spain
\and INAF - Osservatorio Astronomico di Padova, vicolo dell'Osservatorio 5, I-35122 Padova, Italy
\and Instituto de Astronom\'ia, Universidad Nacional Aut\'onoma de M\'exico, Apdo. 106, Ensenada BC 22800, Mexico
}

\abstract
{The intrinsic shape of galactic bulges in nearby galaxies
provides crucial information to separate bulge types.}
{We intended to derive accurate constraints to the 
intrinsic shape of bulges to provide new clues on their
formation mechanisms and set new limitations for future simulations.}
{We retrieved the intrinsic shape of a sample of CALIFA bulges
using a statistical approach. Taking advantage of
\texttt{GalMer} numerical simulations of binary mergers we estimated
the reliability of the procedure. Analyzing the $i$-band mock images
of resulting lenticular remnants, we studied the intrinsic shape of
their bulges at different galaxy inclinations. Finally, we introduced a
new ($B/A$, $C/A$) diagram to analyze possible correlations between
the intrinsic shape and the properties of bulges.}
{We tested the method on simulated lenticular remnants, finding
that for galaxies with inclinations $25^{\circ} \le \theta \le
65^{\circ}$ we can safely derive the intrinsic shape
of their bulges. 
We found that our CALIFA bulges tend to be nearly oblate systems ($66\%$), with
a smaller fraction of prolate spheroids ($19\%$) and triaxial ellipsoids ($15\%$). 
The majority of triaxial bulges are in barred galaxies
($75\%$). Moreover, we found that bulges with low S\'ersic indices or in galaxies with
low bulge-to-total luminosity ratios form a heterogeneous class of objects; additionally, also
bulges in late-type galaxies or in less massive galaxies 
have no preference in being oblate, prolate, or triaxial. On the contrary,
bulges with high S\'ersic index, in early-type galaxies, or in more
massive galaxies are mostly oblate systems.}
{We concluded that various evolutionary pathways may coexist
in galaxies, with merging events and dissipative collapse
being the main mechanisms driving the formation of the most
massive oblate bulges and bar evolution reshaping the less massive
triaxial bulges.}


\maketitle



\section{Introduction \label{sec:introduction}}

\begin{figure*}[t!]
\centering
\includegraphics[width=17cm]{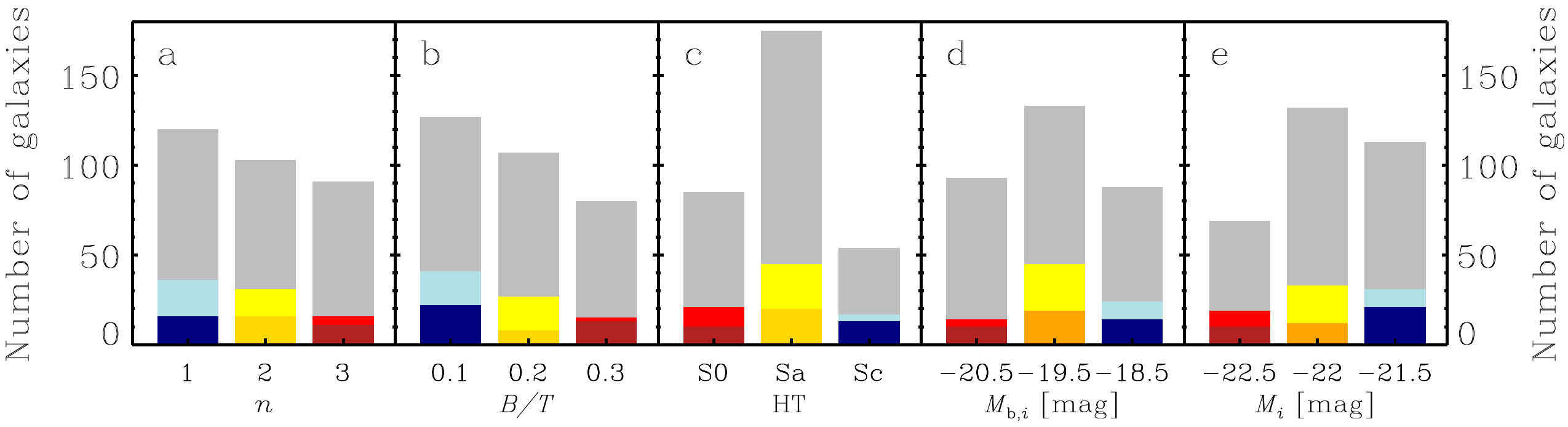}
\caption{Distribution of the S\'ersic index of the bulge (\emph{panel
a}), bulge-to-total luminosity ratio (\emph{panel b}), Hubble type
(\emph{panel c}; Sa bin comprises Sa-Sab-Sb-Sbc galaxies,
while Sc bin comprises Sc-Scd-Sd-Sdm galaxies), 
$i$-band absolute magnitude of the bulge
(\emph{panel d}), and $i$-band absolute magnitude of the galaxy
(\emph{panel e}) for our final sample of 43 unbarred (dark color
histograms) and 40 barred galaxies (light color histograms).  The
grey histograms show the distribution of the remaining 231 galaxies
of the sample of 314 disk galaxies selected from CALIFA DR3. The
galaxy properties are taken from \citet{Mendez2017}.}
\label{fig:sample}
\end{figure*}

In observational extragalactic astrophysics, our measurements of the
light distribution of galaxies are confined to the two-dimensional
framework of the sky plane. Although observations can only access the
projected rather than intrinsic luminosity density of galaxies, we can
disentangle their different luminous components, including
bulges. Constraining the three-dimensional light distribution of the
galaxy components, and therefore their intrinsic shape, is a crucial
piece of information in our understanding of how galaxies form and
evolve.

Several studies addressed the intrinsic shape of the elliptical
galaxies \citep{Sandage1970, Tremblay1996, Rodriguez2013}. Although
many of ellipticals were initially thought to be oblate or prolate spheroids,
some photometric \citep[i.e., the twisting of the
isophotes;][]{Carter1978, Bertola1979} and kinematic properties \citep[i.e., the
low rotation of stars or the kinematic 
misalignment;][]{Bertola1975, Illingworth1977, Krajnovic2011}
promptly supported the idea that some of them
could be triaxial ellipsoids. In general, faint ellipticals are more
flattened with a tendency to be oblate spheroids, whereas bright
ellipticals are rounder and more frequently triaxial
ellipsoids \citep{Weijmans2014}. 
It should be noticed that most of the works about the intrinsic
shape of ellipticals deal with the distribution function of the
intrinsic axial ratios of the whole population of ellipticals through
statistical analyses of their apparent flattenings 
\citep[see][for a review]{Mendez2016}. 
As a matter of fact, it is not possible to recover the intrinsic shape of
an individual elliptical galaxy by only studying its light distribution
\citep{Statler2001}. Indeed, deprojecting the apparent shape of an elliptical
into its intrinsic shape represents a typical ill-posed problem,
caused by the lack of observational constraints on the three Euler
angles ($\theta$, $\phi$, $\psi$) that provide the
transformation. Further details about the galaxy structure, like the
presence of dust lanes, gaseous disks or embedded stellar disks (e.g.,
\object{NGC~5077}, \citealt{Bertola1991b}) or the knowledge of the stellar velocity
field (e.g., \object{NGC~4365}, \citealt{vandenBosch2008}) 
are needed to overcome this problem.

On the contrary, in disk galaxies it is possible to derive the
intrinsic shape of individual bulges because of the presence of the
disk component, whose observed ellipticity provides a proxy for the
bulge inclination, under the assumptions that both the bulge and disk
share the same equatorial plane of symmetry and that disks are
highly-flattened oblate spheroids \citep[but see also][]{Ryden2004,
Ryden2006}. 
  
The bulge is photometrically defined as the structural component
responsible for the light excess measured in the galaxy central
regions, above the inward extrapolation of the exponential
surface-brightness profile of the disk \citep{Andredakis1995,
Balcells2007}. The main concern for bulges is separating their light
contribution from that of the other galaxy components. This is usually
done by means of the photometric decomposition of the galaxy surface
brightness into the contribution of the bulge and disk, and possibly
of a lens, a bar, inner or outer rings, nuclear unresolved 
components, and the spiral arms 
\citep{Peng2002, Laurikainen2005, Gadotti2009, Benitez2013, Erwin2015}. 

Nowadays, it is widely accepted that bulges are not simple
axisymmetric structures in the center of galaxies \citep[see][for a
review]{Mendez2016}. The misalignment between the bulge and disk
isophotes observed in many spirals \citep[e.g., M31,][]{Lindblad1956,
Williams1979} resembles the isophotal twist of ellipticals
and it is similarly interpreted as the signature of bulge
triaxiality. The first quantitative estimate of the triaxiality of
bulges was carried out by \citet{Bertola1991a} by studying the
misalignment between the major axes of the bulge and disk in a sample
of 32 early-type disk galaxies. They found that the mean intrinsic
axial ratio of bulges in the disk plane is $\langle B/A \rangle =
0.86$, while their mean intrinsic flattening in the plane
perpendicular to the disk plane is $\langle C/A \rangle = 0.65$, where
$A$, $B$, and $C$ are the lengths of the semi-axes of the bulge
ellipsoid. This result was later confirmed by \citet{Fathi2003}, who
analyzed the deprojected axial ratio of the galaxy isophotes within
the bulge radius in a sample of 70 disk galaxies, ranging from
lenticulars to late-type spirals. They found $\langle B/A \rangle = 0.79$
and $\langle B/A \rangle = 0.71$ for the bulges in earlier 
and later morphological types, respectively. By
means of a two-dimensional photometric decomposition,
\citet{Mendez2008} measured the structural parameters of both bulges
and disks in a sample of 148 early-to-intermediate spirals, increasing
the statistics of \citet{Bertola1991a} by an order of magnitude. They
found that about $80\%$ of the sample bulges are triaxial ellipsoids
with $\langle B/A \rangle = 0.85$. More recently, \citet{Mendez2010}
introduced a novel statistical method to constrain the intrinsic shape
of individual bulges. The knowledge of the geometric properties
(i.e., the apparent ellipticity and major-axis position angle) of the bulge 
and disk makes it possible to simultaneously compute the probability 
distribution function of the intrinsic axial ratios $B/A$ and $C/A$ for every
single bulge. They revisited the galaxies of the sample of
\citet{Mendez2008} and concluded that $65\%$ of them host oblate 
triaxial bulges while the remaining ones have prolate triaxial
bulges. 

Nevertheless, further efforts are required to better characterize the
intrinsic shape of bulges and, in particular, a higher number
statistics is needed to investigate the correlations between the bulge
shape and galaxy properties. Knowing the intrinsic shape of bulges
completes our understanding of the potential well and orbital
distribution of the stars in the inner regions of galaxies. This will
also help us to explain the origin of the different populations of
classical and disk-like bulges
as well as to address the assembly processes of their host galaxies. 
\citep[e.g.,][]{Athanassoula2005, Brooks2016}.
The general agreement on bulge formation is that rapid dissipative 
collapses \citep{Eggen1962, Sandage1990} or the violent relaxation 
by galactic major merger events \citep{Toomre1977, Kauffmann1996} 
form classical bulges, resembling oblate spheroids 
with intermediate or low flattening and with a certain degree of triaxiality. 
Numerical simulations have also demonstrated the relevance of minor
mergers in the build up of a classical bulge \citep{Aguerri2001,ElicheMoral2006}.
In the context of galaxy formation, cosmological 
simulations highlighted that the actual population of galaxies 
can be represented only by the proper combination of major and minor, 
gas-rich and gas-poor mergers \citep{Oser2012, Naab2014}. Moreover, the recurring coalescence 
of long-lived giant star-forming clumps at high redshift was also proposed 
as a mechanism for the formation of classical bulges
\citep{Dekel2009, Ceverino2015}.
On the contrary, secular processes linked to the 
evolution of galactic substructures (i.e., bars, lenses, ovals, etc.)
reshape the center of galaxies into either boxy/peanut 
components \citep{Erwin2013, Laurikainen2014} or
more flattened disk-like bulges \citep{Kormendy2004, Kormendy2016}. 
In this scenario, disk-like bulges are expected to be axisymmetric
systems, whereas boxy/peanut structures show some degree of
triaxiality \citep{Athanassoula2006}, being the vertically-thick inner parts of bars
resulting from buckling or resonant effects \citep{Combes1981, Lutticke2000}.
Another possible mechanism invoked for disk-like bulge growth is 
the fast disruption of short-lived giant clumps at high redshift, if
no relaxation processes affect the central region of the galaxy
\citep{Hopkins2012, Bournaud2016}.

Currently, the observational separation between classical
and disk-like bulges is usually done by analyzing their observed photometric, 
kinematic, or stellar population properties \citep{Morelli2008, Coelho2011}. 
However, the demarcation lines are often blurred making 
difficult to understand the actual frequency of different bulge types \citep{Costantin2017}. 
Furthermore, bulges can suffer from different processes during their lifetime with some of 
them giving rise to similar observational properties. As a consequence, 
different kind of bulges can coexist in the same galaxy \citep{Athanassoula2005, 
Mendez2014, Erwin2015b}. Nonetheless, the measurements of 
the intrinsic shape of bulges might provide a fundamental additional 
constraint to separate bulge types, as well as limitations for  future 
numerical simulations willing to reproduce realistic galaxies.

In this paper, we analyze the intrinsic shape of the bulges of some of
the disk galaxies observed in the Calar Alto Legacy Integral Field
Area survey Data Release 3 \citep[CALIFA DR3;][]{Sanchez2016}. We aim
at investigating the possible links between the intrinsic shape of
bulges and their observed photometric properties. 
Here, we improve the previous results by \citet{Mendez2008, Mendez2010} by testing the
reliability of their statistical method and setting limits on the
galaxy inclination to its successful application with the help of mock
images of a set of simulated remnant galaxies.
The paper is organized as follows. We present the galaxy sample in
Sect.~\ref{sec:sample}. We summarize the statistical method for
retrieving the intrinsic shape of bulges in Sect.~\ref{sec:method}. We
make use of mock images of simulated remnant galaxies seen at different
inclinations to understand the limits of our analysis in
Sect.~\ref{sec:simulations}. We derive the intrinsic shape of the
bulges of the sample galaxies in Sect.~\ref{sec:shape}. 
We discuss the possible implications of our results for 
galaxy formation in Sect.~\ref{sec:discussion}. 
We summarize our findings in Sect.~\ref{sec:conclusions}.  We adopt $H_0$ = 70 km
s$^{-1}$ Mpc$^{-1}$, $\Omega_{\rm M}$ = 0.3, and $\Omega_{\Lambda}$ =
0.7 as cosmological parameters throughout this work.


\section{Sample selection \label{sec:sample}}

We selected our galaxies sample from the final sample of galaxies
included in the CALIFA DR3, which was drawn from the Sloan Digital Sky
Survey (SDSS) DR7 \citep{Abazajian2009} and comprises 667 nearby
galaxies ($0.005 < z < 0.03$) with an angular isophotal diameter
between 45 and 79.2 arcsec at a surface brightness level of $25$ mag
arcsec$^{-2}$ in the $r$ band.

First, we focussed onto the 314 disk galaxies of CALIFA DR3, 
not interacting or merging, with a
photometric decomposition obtained by \citet{Mendez2017}. 
These galaxies were fitted with either
bulge and disk only (177 galaxies), or with a bar in addition to bulge
and disk (137 galaxies) using the GAlaxy Surface Photometry 2
Dimensional Decomposition code \citep[GASP2D;][]{Mendez2008,
Mendez2014}.

Then, we took into account only the galaxies with good imaging,
that is the absence of either strong fluctuations of the local sky background
around the galaxy or other bright component affecting the photometric
decomposition (e.g., a lens, inner and/or outer rings, and spiral
arms), and all the structural parameters of the bulge, disk, and bar
left free to vary during the fitting process (i.e., the galaxies
flagged as 1,a in Table 1 of \citealt{Mendez2017}). This allowed us to
obtain a subsample of 118
robustly fitted galaxies (67 unbarred and 51 barred galaxies), 
with no bias on the measured structural parameters that
could hamper our analysis of the bulge intrinsic shape. 

\begin{figure}[t!]
\resizebox{\hsize}{!}{\includegraphics{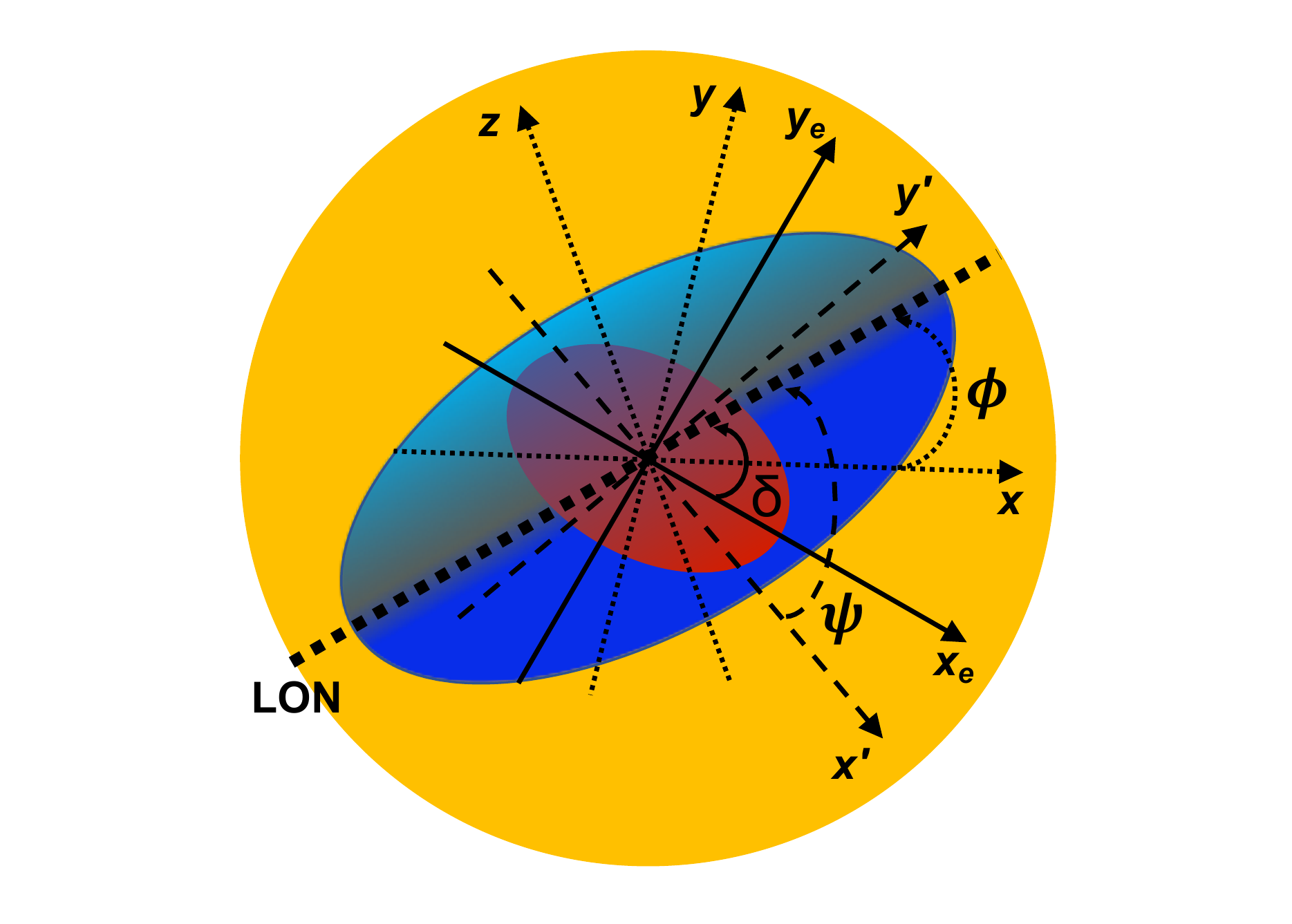}}
\caption{Schematic three-dimensional view of a galaxy with a triaxial
bulge and a infinitesimally thin disk. 
The bulge is shown as seen by the observer
along the LOS. The bulge, disk plane, and sky plane are plotted in
red, blue, and orange, respectively. The reference systems of the
galaxy ($x$, $y$, $z$) and observer ($x'$, $y'$, $z'$) as well as
the LON are plotted with thin dotted lines, thin dashed lines, and a
thick dashed line, respectively. The axes of symmetry ($x_{\rm e}$,
$y_{\rm e}$) of the bulge ellipse in the sky plane are represented
with thin solid lines.}
\label{fig:LOS}
\end{figure}

Finally, we set a limit onto the galaxy inclination ($25^{\circ} <
\theta < 65^{\circ}$) to exclude both the low-inclined galaxies, for
which it is not possible to constrain the bulge shape along the
direction perpendicular to the disk plane, and the highly-inclined
ones, for which the results of the GASP2D photometric decomposition
are not reliable \citep{Mendez2017} 
and the bulge shape on the disk plane is unconstrained
(see Sect. \ref{sec:simulations_4.3} for a discussion). This selection
criterion reduced the galaxy sample to 83 objects 
(43 unbarred and 40 barred galaxies), as shown in 
Table~\ref{tab:appendix:C1} and \ref{tab:appendix:C2}.

We considered only galaxies with $i$-band images to
better resolve the bulge component minimizing the
dust effects with respect to the other SDSS passbands.
The choice of $i$ band assured a sufficient spatial resolution ($FWHM = 1.1 \pm 0.2$ arcsec)
and depth (out to $\mu_i \simeq 26$ mag arcsec$^{-2}$), 
as retrieved from \citet{Mendez2017}.
Their basic properties (i.e., S\'ersic index of the bulge $n$,
bulge-to-total luminosity ratio $B/T$, Hubble type HT, $i$-band
absolute magnitude of the bulge $M_{{\rm b}, \, i}$, and $i$-band
absolute magnitude of the galaxy $M_i$ from \citealt{Mendez2017}) are
shown in Fig.~\ref{fig:sample} and compared with those of the selected
sample of 314 disk galaxies from CALIFA DR3.
This final sample is not complete in volume.
However, we thought that the selection in diameter of the CALIFA sample 
should not introduce any major bias in our results,
since the distribution of bulge observed properties is well sampled, 
as shown in Fig.~\ref{fig:sample}.


\section{Bulge intrinsic shape \label{sec:method}}

The full description of the statistical method adopted to
derive the intrinsic shape of the bulges of our galaxy sample is
given in \citet{Mendez2010}. Here, for the sake of clarity, we
summarized the main hypotheses and gave the most relevant equations
that link the intrinsic axial ratios of the bulge to the observed
properties of the galaxy. 
In particular, we rewrote the description of the
probability function $P(B/A)$ (Eq. 34 in \citealt{Mendez2010}) and we
revised the equation linking the axial ratios $B/A$ and $C/A$ 
(Eq. 59 in \citealt{Mendez2010}).

\begin{table*}
\caption{Intrinsic shape of the bulges of the simulated lenticular remnants seen
at different inclinations.}
\centering
\begin{tabular}{lccccccccc}
\hline
\hline
\noalign{\smallskip}
\multicolumn{1}{c}{Galaxy} & \multicolumn{5}{c}{$B/A$} & & \multicolumn{3}{c}{$C/A$} \\
& ($0^{\circ}$) & ($180^{\circ}$) & ($30^{\circ}$) & ($45^{\circ}$) & ($60^{\circ}$) & &
($30^{\circ}$) & ($45^{\circ}$) & ($60^{\circ}$) \\
\multicolumn{1}{c}{(1)} & (2) & (3) & (4) & (5) & (6) & & (7) & (8) & (9) \\
\noalign{\smallskip}
\hline
\noalign{\smallskip}
gE0gSbo5   & 1.00 & 1.00 & 1.00 & 1.00 & 1.00 & & 0.99 & 0.59 & 0.59\\
gE0gSdo5   & 0.79 & 0.79 & 0.91 & 0.89 & 0.86 & & 0.46 & 0.26 & 0.26\\
gS0dE0o98  & 0.46 & 0.49 & 0.49 & 0.64 & 0.56 & & 0.51 & 0.21 & 0.36\\
gS0dE0o99  & 0.46 & 0.46 & 0.54 & 0.59 & 0.69 & & 0.64 & 0.64 & 0.64\\
gS0dE0o100 & 0.46 & 0.46 & 0.49 & 0.54 & 0.56 & & 0.59 & 0.14 & 0.39\\
gS0dS0o99  & 0.59 & 0.59 & 0.56 & 0.61 & 0.76 & & 0.64 & 0.41 & 0.21\\
gS0dSao103 & 0.51 & 0.51 & 0.51 & 0.56 & 0.56 & & 0.96 & 0.21 & 0.34\\
gS0dSbo106 & 0.44 & 0.44 & 0.41 & 0.46 & 0.61 & & 0.69 & 0.39 & 0.29\\
gS0dSdo100 & 0.51 & 0.51 & 0.49 & 0.51 & 0.74 & & 0.54 & 0.41 & 0.24\\
gSbgSbo9   & 0.71 & 0.71 & 0.76 & 0.76 & 0.76 & & 1.06 & 0.74 & 0.54\\
gSbgSdo5   & 0.81 & 0.84 & 0.84 & 0.81 & 0.84 & & 0.36 & 0.26 & 0.21\\
\noalign{\smallskip}
\hline
\end{tabular}
\tablefoot{(1) Identifier in the \texttt{GalMer} database of the
merger experiment resulting in a lenticular remnant, which we
adopted as the name of the simulated galaxy. (2), (3), (4), (5), (6)
Intrinsic axial ratio $B/A$ of the bulge obtained from the mock
images of the simulated lenticular remnants seen at an inclination $\theta =
0^{\circ}$, $180^{\circ}$, $30^{\circ}$, $45^{\circ}$, and
$60^{\circ}$, respectively. (7), (8), (9) Intrinsic axial ratio
$C/A$ of the bulge obtained from the mock images of the simulated
lenticular remnants seen at an inclination $\theta = 30^{\circ}$, $45^{\circ}$,
and $60^{\circ}$, respectively.}
\label{tab:simulations_shape}
\end{table*}

\subsection{Bulge and disk geometry \label{sec:method_geom}}

In order to characterize the intrinsic shape of a bulge, we assumed it
to be a triaxial ellipsoid with the same equatorial plane as the disk,
which we supposed to be infinitesimally thin. 
Moreover, the bulge and
disk share the same center, which coincides with the galaxy center
(Fig. \ref{fig:LOS}).

Let ($x$, $y$, $z$) be the Cartesian coordinates in the reference
system of the galaxy. The origin of the system is in the galaxy
center, the $x$-axis and $y$-axis correspond to the bulge principal
axes in the equatorial plane, while the $z$-axis corresponds to the
common polar axis of both the bulge and disk. The equation of the
bulge in its own reference system is given by
\begin{equation}
\dfrac{x^2}{A^2} + \dfrac{y^2}{B^2} + \dfrac{z^2}{C^2} = 1 \, ,
\label{eq:ellipsoid}
\end{equation}
where $A$, $B$, and $C$ are the lengths of the bulge intrinsic
semi-axes.

Let ($x'$, $y'$, $z'$) be the Cartesian coordinates in the reference
system of the observer. The origin of the system is in the galaxy
center, the polar $z'$-axis is along the line of sight (LOS) and
points toward the galaxy, while ($x'$, $y'$) confines the sky plane.

The intersection between the bulge equatorial plane ($x$, $y$) and the
sky plane ($x'$, $y'$) is the so-called line of nodes (LON). The angle
$\theta$ between the polar $z$-axis and the polar $z'$-axis defines
the bulge inclination. Let $\phi$ be the angle between the $x$-axis
and the LON in the bulge equatorial plane and let $\psi$ be the angle
between the $x'$-axis and the LON in the sky plane. The three Euler
angles ($\theta$, $\phi$, $\psi$) allow for the transformation from
the reference system of the sky to that of the galaxy. If the
$x'$-axis coincides with the LON, consequently it is $\psi=0$.

The projection onto the sky plane of the triaxial ellipsoid given in
Eq.~\ref{eq:ellipsoid} is an ellipse which corresponds on the
galaxy image to the photometric bulge. It is given by
\begin{equation}
\dfrac{x_{\rm e}^2}{a^2} + \dfrac{y_{\rm e}^2}{b^2} = 1 \, ,
\end{equation}
where $x_{\rm e}$ and $y_{\rm e}$ are taken along the symmetry axes of
the bulge ellipse, while $a$ and $b$ are lengths of the ellipse
semi-major and semi-minor axis, respectively. The twist angle $\delta$
between the $x_{\rm e}$-axis and the LON indicates the bulge
orientation. We always consider $0 < \delta < 90^{\circ}$ 
such that $a$ can be either the major or the minor semi-axis.

The twist angle $\delta$ and apparent axial ratio of the bulge
($q_{\rm b} = b/a$) depend only, and unambiguously, on the direction
of the LOS (i.e., on $\theta$, $\phi$, and $\psi$) and on the
intrinsic shape of the bulge (i.e., on $A$, $B$, and $C$).  Thus, by
means of the Euler angles it is possible to impose further constraints
to the equations that relate the intrinsic parameters of the bulge in
the reference system of the galaxy to the observed properties of the
galaxy in the reference system of the observer \citep{Simonneau1998,
Mendez2008}. Introducing physical constraints on the accessible
viewing angles (e.g., imposing that $A$, $B$, and $C$ must be definite
positive), it is possible to statistically derive the intrinsic axial
ratios of the bulge from its observed properties. Unfortunately, the
problem is not analytically solved because of the unknown
spatial position of the
bulge (i.e., the angle $\phi$), which constitutes the basis of the
statistical analysis \citep[see][for all the details]{Mendez2010}.

\subsection{Statistical analysis \label{sec:method_stat}}

The theoretical framework based on the statistical analysis of the
$\phi$ angle allows us to retrieve the intrinsic shape of individual
bulges in disk galaxies from the bulge apparent shape (i.e., the bulge
ellipticity $\epsilon_{\rm b} = 1- q_{\rm b}$), the disk apparent
shape (i.e., the disk ellipticity $\epsilon_{\rm d} = 1- q_{\rm d}$,
where $q_{\rm d}$ is the apparent axial ratio of the disk), and the
bulge twist angle (i.e., the difference between the position angles of
the bulge and disk $\delta = PA_{\rm d} - PA_{\rm b}$;  
see \citealt{Mendez2010} for all details). 

The apparent axial ratio of a circular and infinitesimally thin 
disk is a measure of the bulge inclination
\begin{equation}
\theta = \arccos{q_{\rm d}} \, ,
\end{equation}
when both the bulge and disk share the same equatorial plane.
However, disks are not infinitely thin structures \citep{Sandage1970,
Ryden2004}. To account for this, we computed the inclination of our
galaxies accounting for the distribution function of the intrinsic
axial ratio of the disks $q_{0, \rm d}$. We adopted a normal
distribution function with mean intrinsic axial ratio $\langle q_{0,
\rm d} \rangle = 0.267$ and standard deviation $\sigma_{q_{0, \rm
d}} = 0.102$ following \citet{Rodriguez2013}. Thus, the
statistical value of the galaxy inclination is
\begin{equation}
\theta = \arccos \sqrt{\dfrac{q_{\rm d}^2 - q_{0, \rm d}^2}{1 -q_{0, \rm d}^2}} \, ,
\end{equation}
where the $q_{0, \rm d}$ value is randomly drawn from the previous
normal distribution.

We took into account for the uncertainties in $\epsilon_{\rm b}$,
$\epsilon_{\rm d}$, and $\delta$ derived from the error analysis of
the photometric decomposition \citep{Mendez2017}. We randomly
generated 1000 geometric configurations by adopting for each parameter
a Gaussian distribution centered on its measured value and with a
standard deviation equal to its uncertainty. A similar analysis was
introduced by \citet{Corsini2012} to recover the intrinsic shape of
the polar bulge of \object{NGC~4698}. For each geometric configuration, we
calculated 5000 values of $B/A$ using Monte Carlo simulations and
according to its probability function
\begin{equation}
P\left(\dfrac{B}{A}\right) = \dfrac{2 \, \dfrac{B}{A} \sin\phi_B}{(\phi_C - \phi_B) \left(1 - \dfrac{B^2}{A^2} \right) \sqrt{\left(1 - \dfrac{B^2}{A^2} \right)^2 - \sin^2{\phi_B} \left(1 + \dfrac{B^2}{A^2} \right)^2}} \, ,
\label{eq:BA}
\end{equation}
where $\phi_B$ and
$\phi_C$ are the angles where the length of the intrinsic semi-axis
$B$ and $C$ are zero, respectively.
Since $B/A$ and $C/A$ are both functions of the same variable
$\phi$, their probabilities $P(B/A)$ and $P(C/A)$ are equivalent;
thus, after sampling the value of $B/A$ using Eq.~\ref{eq:BA}, we
calculated the value of $C/A$ using
\begin{equation}
\begin{split}
\dfrac{2 \sin(2 \phi_C)}{F_{\theta}} \dfrac{C^2}{A^2} = & \sin(2\phi_C - \phi_B) \sqrt{\left(1-\dfrac{B^2}{A^2}\right)^2 - \sin^2\phi_B \left(1 + \dfrac{B^2}{A^2}\right)^2} \\
& - \sin\phi_B\cos(2\phi_C - \phi_B) \left(1 + \dfrac{B^2}{A^2} \right) \, ,
\end{split}
\label{eq:probability}
\end{equation}
where $\phi_B$, $\phi_C$, and $F_{\theta}$ are functions of the
observed quantities $a$, $b$, $\delta$, and $\theta$ \citep[see][for a
full description of the different variables]{Mendez2010}.


\section{Bulge intrinsic shape of the simulated lenticular remnants \label{sec:simulations}}

\subsection{Simulated lenticular remnants from numerical experiments of binary mergers}

In order to test our statistical method for recovering the intrinsic
shape of bulges and to understand its limitations, we measured the
intrinsic axial ratios of the bulge against different galaxy
inclinations. To this aim, we used a subset of 11 numerical
simulations from the \texttt{GalMer} database\footnote{The
\texttt{GalMer} database is a public library of hydrodynamics N-body
simulations of galaxy mergers with intermediate resolution available
at \url{http://www.project-horizon.fr/}.}  \citep{Chilingarian2010}.
We used remnant galaxies of a variety of merger experiments
between pairs of galaxies with different mass, morphology, gas
content, and orbital parameters. These remnant galaxies strongly resemble
lenticular galaxies, according to their morphological, photometric,
and kinematic properties \citep[][Eliche-Moral et al. in prep.]{Borlaff2014, Querejeta2015a,
Querejeta2015b, Tapia2017}.

We chose to analyze simulated lenticular remnants resulting from binary mergers
instead of N-body realizations of analytical expressions, as those
adopted for building the progenitor galaxies of the remnants, because
we required a certain degree of bulge triaxiality. Such a triaxiality
is a common feature of merger remnants, although their progenitors
could be axisymmetric by construction \citep{Cox2006, Tapia2014}. 
All the simulated lenticular remnants of the analyzed merger experiments are
either unbarred galaxies if resulting from a major merger, or weakly
barred galaxies if resulting from a minor merger (Eliche-Moral et al. in prep.).
We discarded the merger experiments producing elliptical or E/S0 remnants, in order
to have simulated galaxies with a well-defined bulge embedded into a
large disk, similarly to the observed CALIFA galaxies.

\begin{table}
\caption{Orbital parameters of the merger experiments resulting in the
simulated lenticular remnants listed in Table~\ref{tab:simulations_shape}.}
\centering
\begin{tabular}{ccccc}
\hline
\hline
\noalign{\smallskip}
ID$_{\rm orb}$ & spin-orbit & $i_2$     & $d_{\rm per}$ & $E_0$ \\
             & [P/R]      & [$^\circ$] & [kpc]       & [$10^4$ km$^2$ s$^{-2}$] \\
(1) & (2) & (3) & (4) & (5) \\
\noalign{\smallskip}
\hline
\noalign{\smallskip}
5   & P   & 0   & 16  & 0   \\
9   & P   & 0   & 24  & 0   \\
98  & P   & 33  &  8  & 2.5 \\
99  & P   & 33  &  8  & 5   \\
100 & P   & 33  &  8  & 15  \\
103 & R   & 33  &  8  & 0   \\
106 & R   & 33  &  8  & 15  \\
\noalign{\smallskip}
\hline
\end{tabular}
\tablefoot{(1) Identifier in the \texttt{GalMer} database of the orbit
used in the merger experiment. (2) Spin-orbit coupling (P: prograde;
R: retrograde). 
(3) Inclination of the
secondary progenitor with respect to the orbital plane. 
(4) Pericenter distance. (5) Initial orbital
energy.}
\label{tab:simulations_param}
\end{table}

\begin{figure*}[t!]
\centering
\includegraphics[width=17cm]{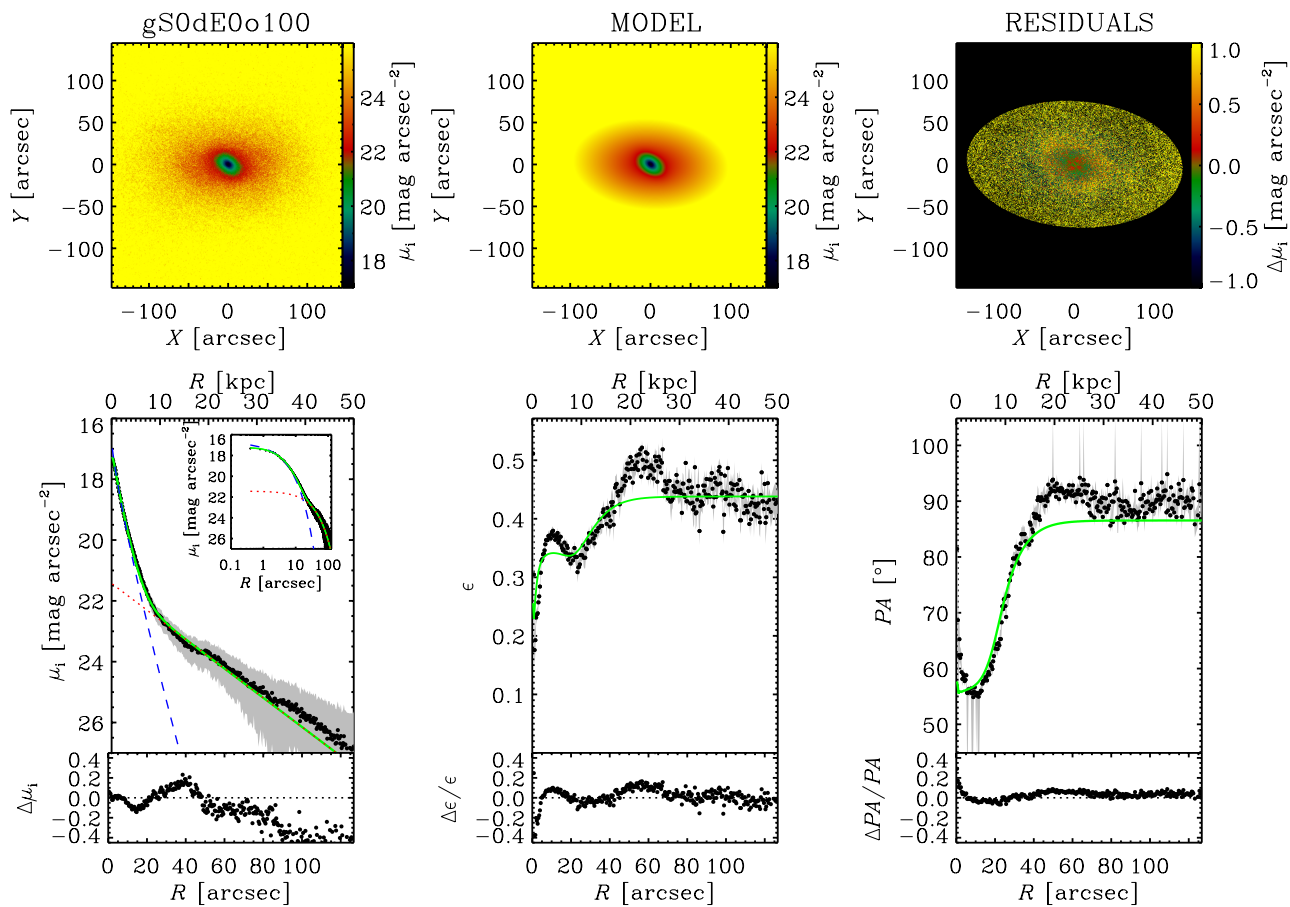}
\caption{Two-dimensional photometric decomposition of the mock
$i$-band image of the simulated lenticular remnant gS0dE0o100 seen at an
inclination $\theta = 60^{\circ}$ obtained with GASP2D. The \emph
{upper panels} (\emph{from left to right}) show the map of the
observed, modeled, and residual (observed$-$modeled) surface
brightness distributions. The \emph{lower panels} (\emph{from left
to right}) show the ellipse-averaged radial profile of surface
brightness, ellipticity, and position angle measured in the observed
(black dots with grey error bars) and seeing-convolved modeled image
(green solid line) and their corresponding difference. The intrinsic
surface-brightness radial profiles of the best-fitting bulge (blue
dashed line) and disk (red dotted line) are also shown in both
linear and logarithmic scale for the distance to the center of the
galaxy.}
\label{fig:simulations_result}
\end{figure*}

The progenitor galaxies were modeled with a spherical non-rotating
dark-matter halo, which contains a stellar and/or a gaseous disk
and/or a central non-rotating bulge, depending on their morphological
type.  The primary galaxy consisted of a giant galaxy (hereafter gE0
for a giant-like elliptical, gS0 for a giant-like lenticular, gSa for
a giant-like Sa spiral, gSb for a giant-like Sb spiral, and gSd for a
giant-like Sd spiral) interacting either with another giant galaxy of similar mass or
with a dwarf galaxy (hereafter dE0, dS0, dSa, dSb, and dSd), whose
total mass is ten times smaller than that of the giant galaxy.
Several simulations were performed varying the initial orbital energy,
pericenter distance and inclination with respect to the orbital plane
of the interacting galaxies. 
Indeed, for each interacting pair the disk (when present) 
of one of the two galaxies is kept in the orbital plane, while
the companion disk can have a different inclination.
Direct and retrograde orbits were also
taken into account, where direct or retrograde spin-orbit coupling refer 
to progenitors having either parallel or antiparallel spins, respectively.
The merger experiments have a total of 240\,000
and 528\,000 particles for the major and minor merger events,
respectively. The particles have a mass $M = 3.5$--$20.0 \times 10^5$
M$_{\sun}$ each.
The merger experiments have a duration of 3--3.5 Gyr and were evolved
using a \texttt{Tree-SPH} code \citep{Semelin2002}, 
adopting the same softening length for all
particle types $\epsilon = 280$ pc for the giant-giant galaxy mergers
and $\epsilon = 200$ pc for the giant-dwarf galaxy mergers. The effects
of gas and star formation (SF; such as the stellar mass loss, metallicity
enrichment of the interstellar medium, and energy injection due to
supernova explosions) were considered using the method described in
\citet{Mihos1994}.
The stellar mass of the lenticular remnants is in the range $M =
1$--$3 \times 10^{11}$ M$_{\sun}$ for major mergers and $M =
1.2$--$1.3 \times 10^{11}$ M$_{\sun}$ for minor ones.

The merger experiments we analyzed were chosen to cover the whole
range of morphologies, mass ratios, and orbital configurations of the
progenitors. They are listed in Table~\ref{tab:simulations_shape} and
labelled considering both the morphological type of the progenitors
and the unique numerical identifier given to the orbit in the
\texttt{GalMer} database. The orbital configuration of each merger
experiment is provided in Table~\ref{tab:simulations_param}. For
example, the experiment gS0dE0o100 corresponds to the accretion of a
dwarf elliptical by a giant-like lenticular. It follows the orbit tagged
as 100 in the \texttt{GalMer} database with an inclination of
$33^{\circ}$ with respect to the orbital plane, a pericenter distance
of 8 kpc, and a initial energy of $15 \times 10^4$ km$^2$ s$^{-2}$ in
a prograde spin-orbit coupling. 

\subsection{Photometric decomposition of the simulated lenticular remnants
seen at different inclinations}

To perform a fair comparison between the results from our sample 
of simulated bulges and the final observed sample from CALIFA, we built 
mock images of the simulated lenticular remnants under the observing 
setup of the CALIFA galaxies.

Therefore, we mimicked SDSS $i$-band images of the simulated lenticular remnants
assuming they are at a distance of 67 Mpc, which corresponds to the
median distance of the CALIFA DR3 galaxies. 
We modeled the point spread function (PSF) with a circular Moffat
profile \citep{Moffat1969} with $FWHM = 1.2$ arcsec and $\beta = 5$,
which represent typical values for the SDSS images of the galaxies in
CALIFA DR3 \citep{Mendez2017}.
Moreover, we considered a Poissonian
photon noise to yield signal-to-noise ratio $S/N = 1$ at a limiting
magnitude of $\mu_i = 25.7$ mag arcsec$^{-2}$.
We chose a pixel scale of 0.396 arcsec
pixel$^{-1}$ and, for simplicity, we assumed a gain of $1$ $e{^-}$
ADU$^{-1}$ and a readout noise of 1 $e{^-}$ rms.

We converted the mass of each particle of the simulated lenticular remnants into
light by adopting the $i$-band mass-to-light ratio $M/L$ corresponding
to the stellar population of the same age and metallicity of the
particle. For the old stellar particles, we assumed 
that they have evolved previous to the merger following
a typical star formation history (SFH), according to
the morphological type of the progenitors as found in real galaxies
\citep{ElicheMoral2010}. Since the SF is transferred to the hybrid particles
at the start of the merger simulation, the SFH of 
the old stellar particles is stopped at that moment and 
they are assumed to evolve passively since then. 
We have thus adopted a present-day age of 11 Gyr for the old
stellar component because it is the average age of the old stellar population in
the disks of nearby lenticular galaxies \citep{Silchenko2012,
Silchenko2013}. 
The SFHs were estimated using the stellar population synthesis models by
\citet{Bruzual2003} with a Chabrier initial mass function
\citep{Chabrier2003}, and the
evolutionary tracks by \citet{Bertelli1994}. 
Concerning the hybrid particles, the SF in the galaxies that merge 
is transferred to them during the simulation. So, part of their initial mass
(totally gaseous at the start of the simulation) turns into stellar 
mass during the merger depending on the local gas concentration. 
The SFH of these particles is specifically computed during the experiment 
and it is different for each particle \citep{Chilingarian2010}. 
Although it may be quite complex, most of their SF accumulates 
into one or two short peaks occurred soon after the first pericenter passage 
and the full merger, mostly in this last one 
\citep[see][and Eliche-Moral et al., in prep.]{DiMatteo2007, DiMatteo2008, Lotz2008}. 
Therefore, we have approximated the complex 
SFH of each hybrid particle by simple stellar populations (SSPs), 
assuming the mean age and metallicity that each hybrid particle 
presents at the end of the simulation, to estimate a $M/L$ ratio for each 
one and convert their newly formed stellar mass into luminosity. 
For this goal, we have used the same stellar population synthesis models commented before.
We transformed the intrinsic physical values of lengths into projected angular values and
we corrected the resulting surface brightness by cosmological dimming
\citep[see][for more details]{Tapia2017}.

\begin{figure}[t!]
\centering
\resizebox{\hsize}{!}{\includegraphics{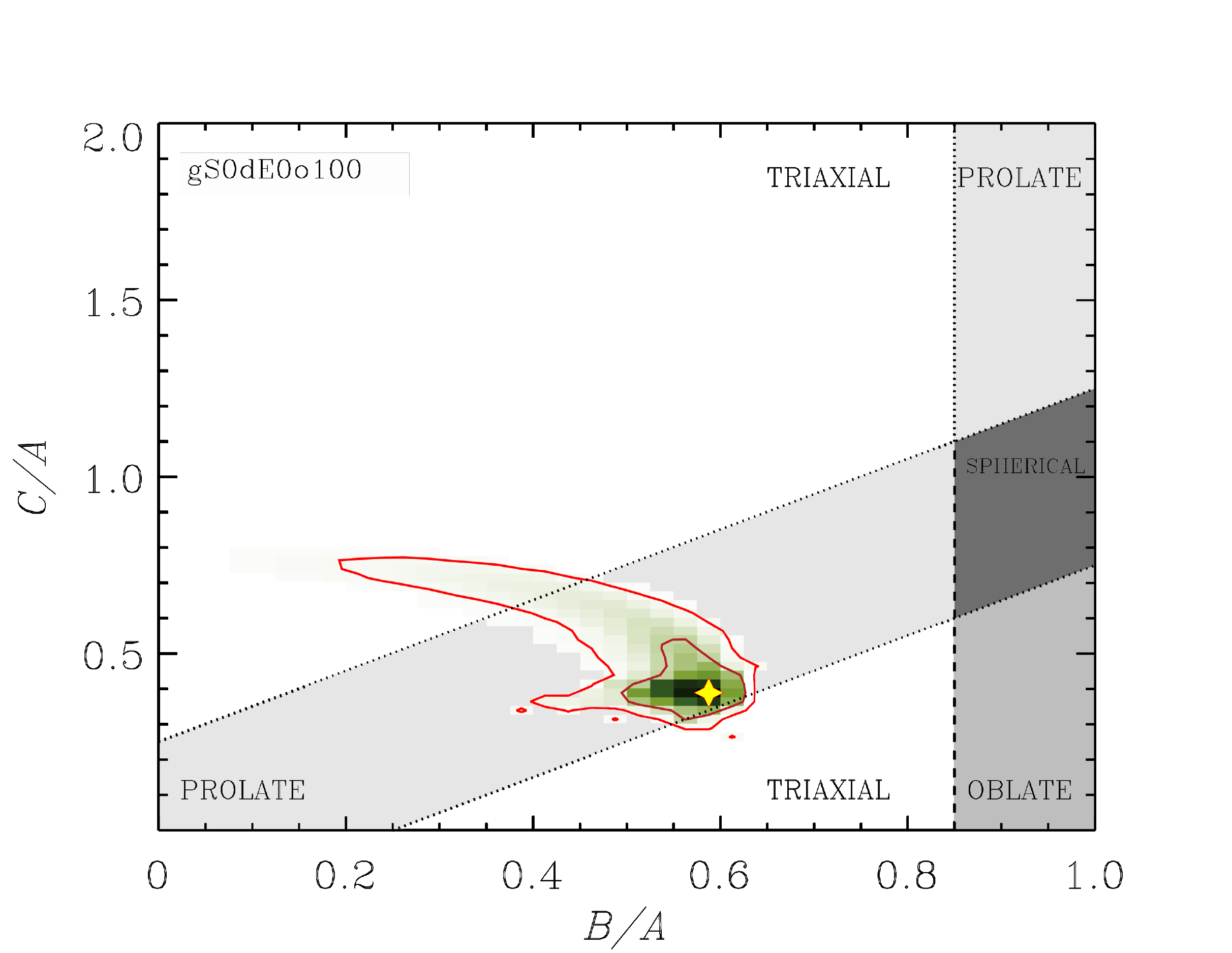}}
\caption{Distribution of the intrinsic axial ratios $B/A$ and $C/A$ of
the bulge of the simulated lenticular remnant gS0dE0o100 seen at an inclination
$\theta = 60^{\circ}$. The yellow star corresponds to the most
probable values of $B/A$ and $C/A$. The inner and outer red solid
contours encompass respectively the $68.3\%$ and $95.4\%$ of the
realizations of $(B/A,C/A)$ consistent with the geometric parameters
of bulge and disk measured from our photometric decomposition of the
mock image of the simulated lenticular remnant. 
The white, light grey, grey, and dark grey regions
mark the regimes of triaxial, prolate, oblate, and spherical bulges,
respectively. }
\label{fig:shape_sim_single}
\end{figure}

For each simulated lenticular remnant, we created 6 mock images corresponding to
different inclinations with respect to the direction of the total
angular momentum vector of the simulated lenticular remnant (i.e., $\theta =
0^{\circ}$, $30^{\circ}$, $45^{\circ}$, $60^{\circ}$, $90^{\circ}$,
and $180^{\circ}$). The face-on views
$\theta = 0^{\circ}$ and $180^{\circ}$ correspond to the cases where
the angular momentum vector points towards to and away from the
observer, respectively. This allowed us to compare the reliability of
the photometric decomposition results of both cases (which should be
identical) and the dependence of our method to derive the intrinsic
shape of bulges on the galaxy inclination.  We analyzed the mock
images of the simulated lenticular remnants as if they were real by performing a
photometric decomposition with GASP2D.  We modeled the surface
brightness of the bulge with a S\'ersic law \citep{Sersic1968}, the
surface brightness of the disk either with a single exponential
\citep{Freeman1970} or with a double-exponential law
\citep{vanderKruit1979}, and the surface brightness of the bar with a
Ferrers law \citep{Ferrers1877, Aguerri2009}. 
An example of the GASP2D photometric decomposition of the mock images
of the simulated galaxies is shown in
Fig.~\ref{fig:simulations_result} for the lenticular remnant resulting
from the merger experiment gS0dE0o100 seen at an inclination
$\theta=60^\circ$. 

\begin{figure}[t!]
\centering
\resizebox{\hsize}{!}{\includegraphics{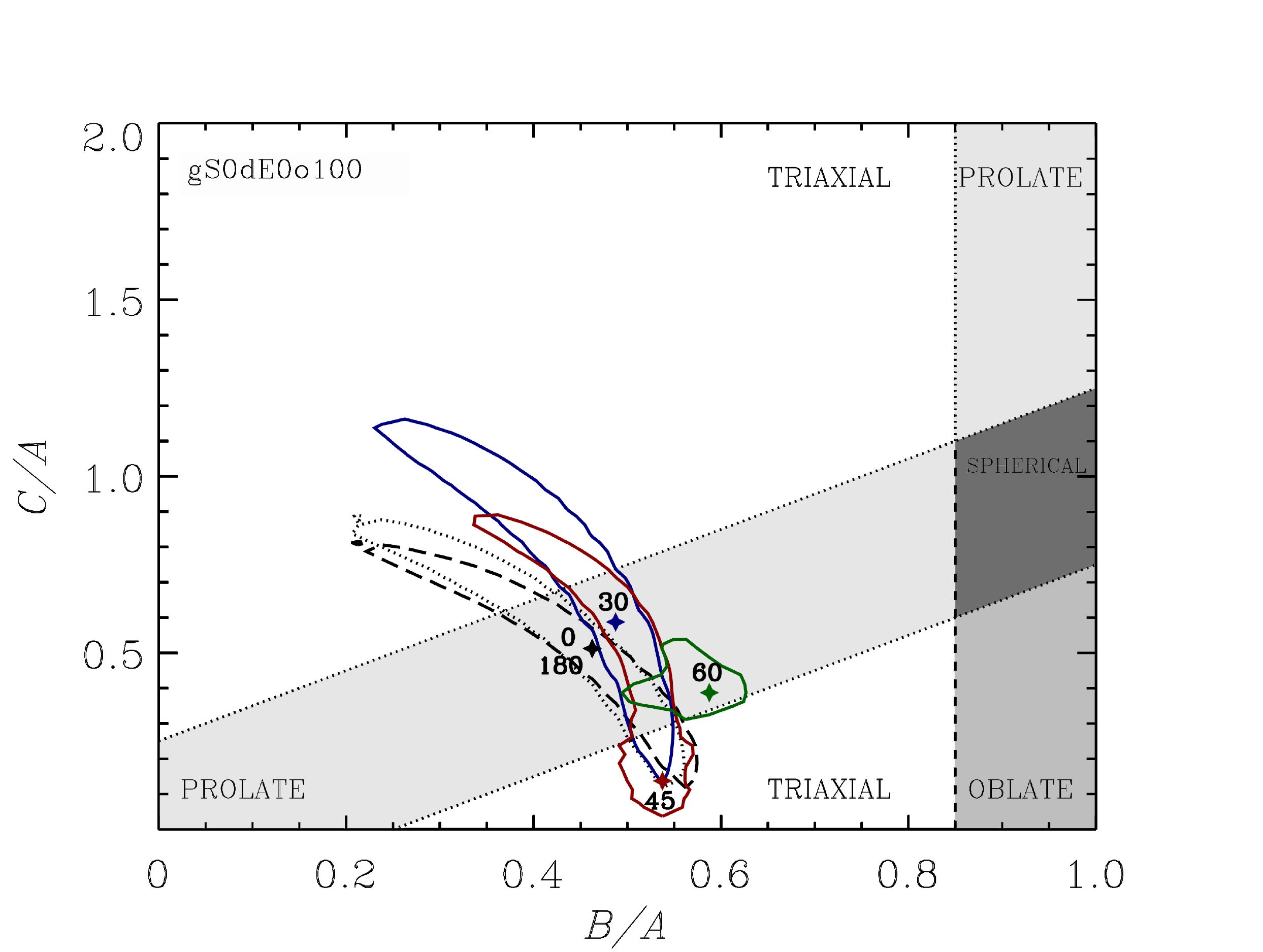}}
\caption{Distribution of the intrinsic axial ratios $B/A$ and $C/A$ of
the bulge of the simulated lenticular remnant gS0dE0o100 seen at different
inclinations. The contours encompass the $68.3\%$ of the
realizations of $(B/A,C/A)$ consistent with the geometric parameters
of bulge and disk measured from our photometric decomposition of the
mock images of the simulated lenticular remnant at $\theta = 0^{\circ}$ (black dashed line),
$30^{\circ}$ (blue), $45^{\circ}$ (red), $60^{\circ}$ (green), and
$180^\circ$ (black dotted line).
The stars correspond to the most probable values of $B/A$ and $C/A$ for the
different galaxy inclinations and are color coded as their
corresponding contours. The white, light grey, grey, and dark grey regions
mark the regimes of triaxial, prolate, oblate, and spherical bulges,
respectively.}
\label{fig:simulations_shape}
\end{figure}

We listed the more relevant best-fitting structural parameters (i.e.,
the effective radius $r_{\rm e}$, S\'ersic index $n$, and axial ratio
$q_{\rm b}$ of the bulge, the scale-length $h$ and axial ratio $q_{\rm
d}$ of the disk, and the difference between the position angles of
bulge and disk $\delta$) of the mock images of the simulated lenticular remnants
seen at different inclinations in Table \ref{tab:photometry_1} for the
giant-giant galaxy mergers, Table \ref{tab:photometry_2} for the
giant-dwarf galaxy mergers, and Table \ref{tab:photometry_3} for the
giant S0-dwarf E0 galaxy mergers with different orbital
parameters. Following \citet{Mendez2017}, we adopted
$\sigma_q = 0.01$ and $\sigma_{PA} = 1^{\circ}$ as uncertainties on
the axial ratio and position angle of both bulge and disk,
respectively.

\subsection{Bulge intrinsic shape of the simulated lenticular remnants seen
at different inclinations \label{sec:simulations_4.3}}

We made use of our statistical method to retrieve from the mock images
the probability distribution of the intrinsic axial ratios $B/A$ and
$C/A$ of the bulges of the simulated lenticular remnants seen at different
inclinations ($\theta = 0^\circ, 30^\circ, 45^\circ, 60^\circ$, and
$180^\circ$). We excluded from
the analysis the edge-on configurations ($\theta = 90^\circ$) because
they do not allow us to constrain $B/A$ and $C/A$ due to the unknown
orientation of the triaxial bulge in the disk plane.
The probability distribution of $B/A$ and $C/A$ for the bulge of the
simulated lenticular remnants resulting from the merger experiment gS0dE0o100 and
seen at an inclination $\theta=60^\circ$ is shown as an example in
Fig. \ref{fig:shape_sim_single}.

The face-on configurations of the simulated lenticular remnants ($\theta =
0^\circ$ and $180^\circ$) provided the same result in terms of the
probability distribution of $B/A$ and $C/A$ for all the simulated
bulges, as expected if no observational and theoretical bias affected
the adopted statistical method. Furthermore, we confirmed that the
tightest constraints for $B/A$ are given when galaxies are seen face
on, whereas $C/A$ remains unconstrained for these galaxies.
We also found consistent probability distributions of $B/A$ and $C/A$
for the same simulated lenticular remnant seen at intermediate inclinations
($\theta = 30^{\circ}$, $45^{\circ}$, and $60^{\circ}$). This made us
confident of having correctly recovered the bulge intrinsic shape and
suggested us to set a limit on the inclination of real galaxies
($25^{\circ} < \theta < 65^{\circ}$) to robustly apply our
statistical method to their bulges (see Sect. \ref{sec:sample}). In
general, the probability distribution of $B/A$ and $C/A$ is tighter at
$\theta = 60^{\circ}$ with respect to $\theta = 30^{\circ}$ or
$45^{\circ}$. Therefore, we considered this inclination as the ideal
viewing angle for future analyses of the bulge intrinsic shape in real
galaxies.

The probability distributions of $B/A$ and $C/A$ of the bulge of the
simulated lenticular remnant resulting from the merger experiment gS0dE0o100 seen
at different inclinations are shown in Fig.~\ref{fig:simulations_shape}, 
while the remaining galaxies are in Appendix \ref{appendix:B} (Fig. \ref{fig:simulations_shape_B}). 
The values of $B/A$ and $C/A$ derived for all the simulated lenticular remnants seen at
different inclinations are listed in
Table~\ref{tab:simulations_shape}.

We realized that the accuracy of the photometric decomposition is
critical to successfully constrain the bulge intrinsic shape. 
We let free to vary all the
structural parameters of the bulge, disk, and bar in the photometric
decomposition of the mock images with GASP2D.
For a few galaxies, we found that the $1\sigma$ level contours of the
probability distributions of $B/A$ and $C/A$ obtained at different
inclinations did not overlap. We double checked the photometric
decomposition of these galaxies and noticed that the ellipticity
and/or position angle of their disks were not well fitted by the
model. As a matter of fact, we fitted the surface brightness
distribution of all these disks with a double-exponential profile and
assumed they had the same ellipticity and position angle both in the
inner and outer regions.
We found that in some cases (e.g., gE0gSbo5) the change in the
ellipticity and position angle measured at the break radius was
probably due to the fact that there were two distinct structural
components (i.e., a lens and a disk) with different geometrical
parameters, instead of a single down- or up-bending exponential
disk. As a consequence, the adopted photometric model did not
exquisitely match the surface brightness distribution of the simulated
lenticular remnant. In other cases (e.g., gSbgSdo5), the change was due a moderate
degree of granularity observed in the mock images at large
galactocentric distances caused by light spots coming from
isolated group of stellar particles orbiting the galaxy outskirts.
To address these issues, we refined the estimate of $\epsilon_{\rm d}$
and $PA_{\rm d}$ by assuming the average ellipticity and position
angle of the galaxy isophotes fitted at large radii with the
IRAF\footnote{Image Reduction and Analysis Facility is distributed by
the National Optical Astronomy Observatory (NOAO), which is operated
by the Association of Universities for Research in Astronomy (AURA),
Inc. under cooperative agreement with the National Science
Foundation.}  task \texttt{ellipse} (nominal values
in Table \ref{tab:photometry_1}).


\section{Bulge intrinsic shape of our CALIFA galaxies \label{sec:shape}}

\begin{figure}[t!]
\centering
\resizebox{\hsize}{!}{\includegraphics{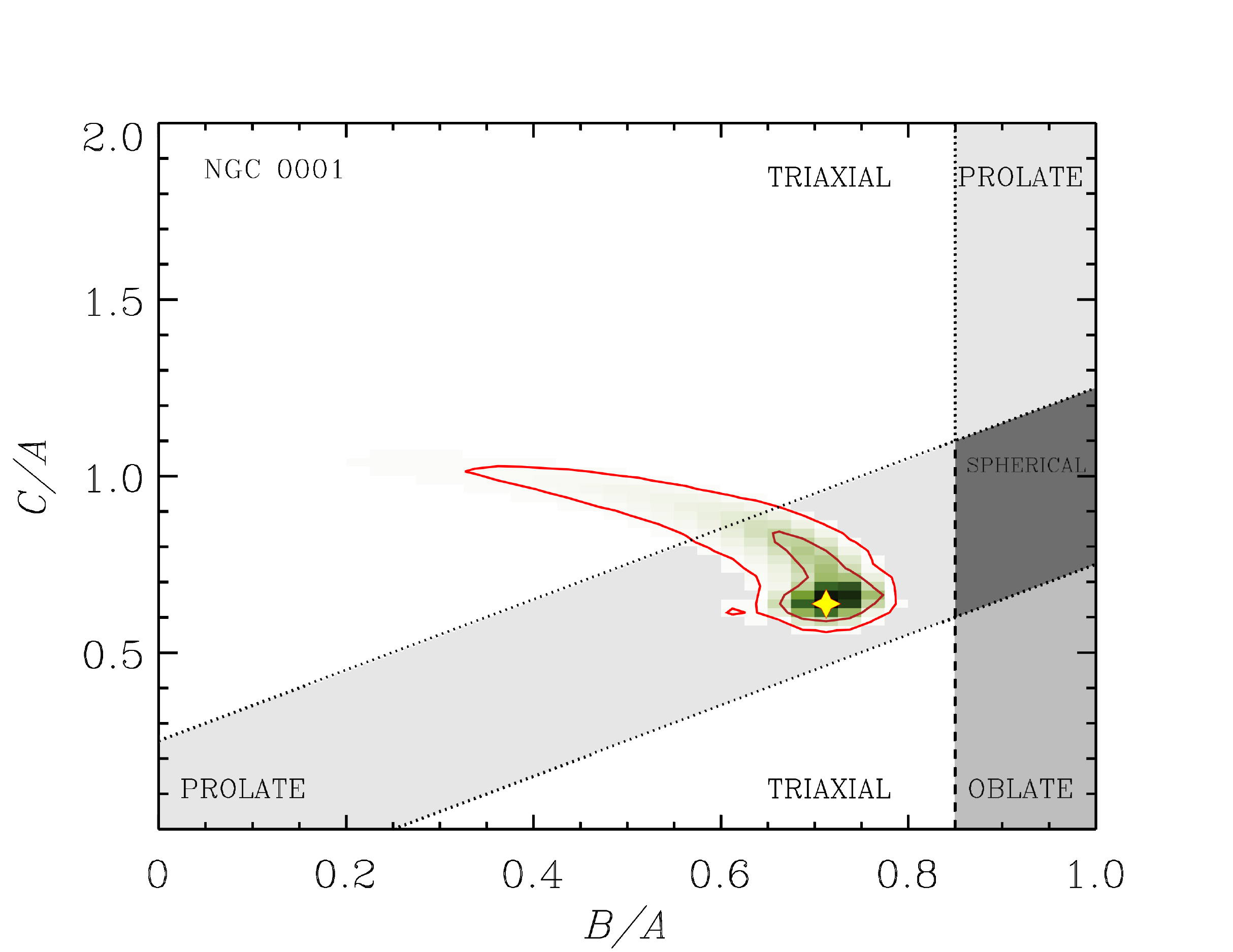}}
\caption{As in Fig.~\ref{fig:shape_sim_single}, but for the bulge of
\object{NGC~1}. The inner and outer red solid contours encompass respectively
the $68.3\%$ and $95.4\%$ of the realizations of $(B/A,C/A)$
consistent with the geometric parameters of bulge and disk measured
by \citet{Mendez2017} with a photometric decomposition of the SDSS
$i$-band image of the galaxy.}
\label{fig:shape_single}
\end{figure}

\begin{figure*}[t!]
\includegraphics[width=17cm]{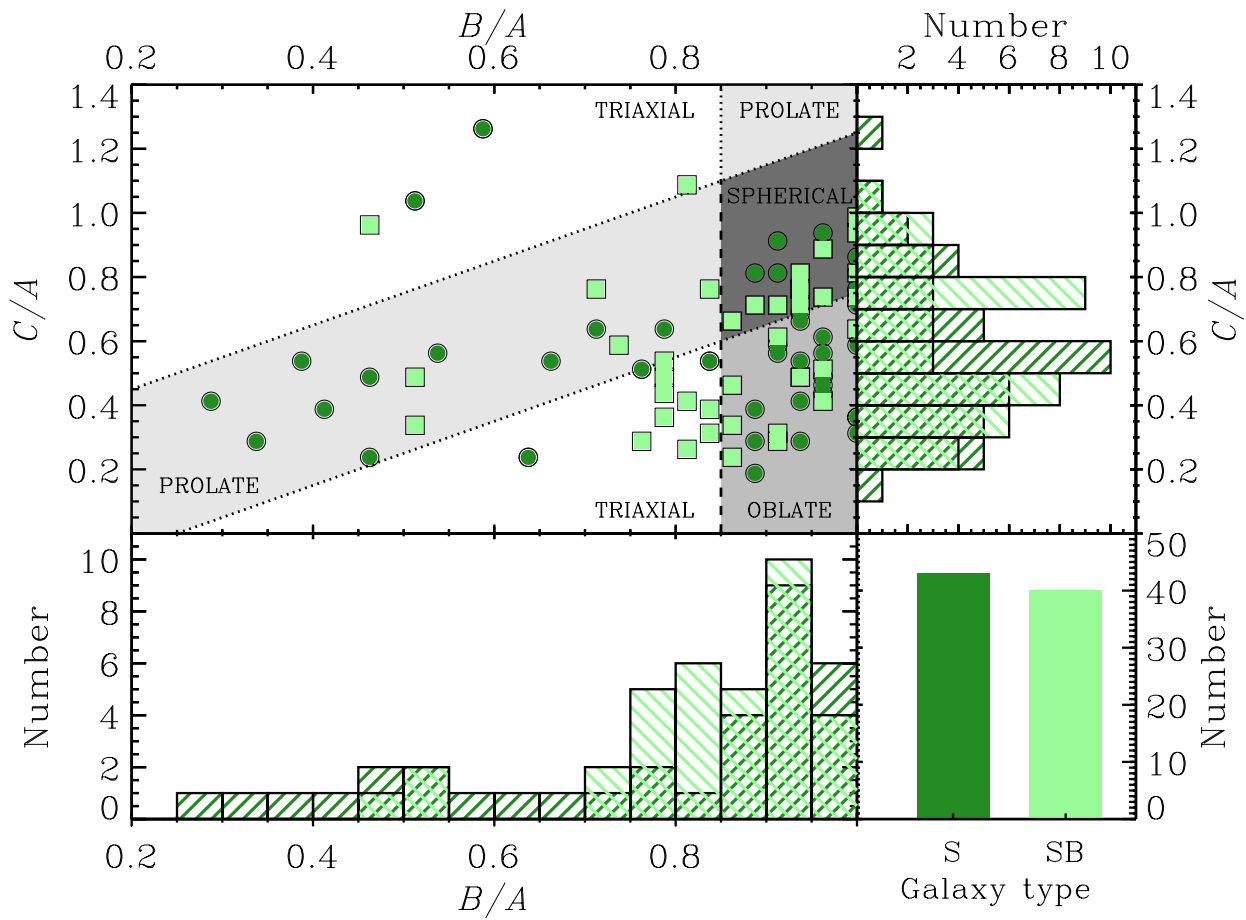}
\caption{\emph{Top left panel:} Intrinsic axial ratios $B/A$ and $C/A$
of our CALIFA bulges. Dark green circles and light green squares 
correspond to unbarred and barred galaxies, respectively. 
The white, light grey, grey, and dark grey regions mark the regimes of triaxial, 
prolate, oblate, and spherical bulges, respectively.
 \emph{Top right panel:} Distribution of $C/A$.
\emph{Bottom left panel:} Distribution of $B/A$.
\emph{Bottom right panel:} Distribution of our CALIFA bulges in
unbarred (S) and barred galaxies (SB).}
\label{fig:shape_tot}
\end{figure*}

We made use of our statistical method to retrieve the probability
distribution of $B/A$ and $C/A$ of the bulges of our CALIFA galaxy
sample. The probability distribution of $B/A$ and $C/A$ for the bulge
of \object{NGC~1} is shown as an example in Fig. \ref{fig:shape_single}, while the most
probable values of $B/A$ and $C/A$ of our CALIFA bulges in
Fig. \ref{fig:shape_tot} and Table~\ref{tab:appendix:C1} and \ref{tab:appendix:C2}. 

We derived for each of our CALIFA bulges the projection of the $1\sigma$
contour along the $B/A$ and $C/A$ axes and adopted the median values
of such projections as the uncertainties on the derived values of
$B/A$ and $C/A$. We estimated $\sigma_{B/A} = 0.15$ and
$\sigma_{C/A} = 0.25$, respectively. At this point, we considered as
oblate spheroids all the bulges with $B/A > 0.85$ 
and $C/A < B/A - 0.25$ (oblate in-plane) or with 
$B/A$ < 0.85 and 0.75 < $C/A$ < 1.25 (oblate off-plane), as prolate
spheroids all the bulges with both $B/A < 0.85$ and $B/A - 0.25 < C/A
< B/A + 0.25$ (prolate in-plane) 
or with $B/A > 0.85$ and $C/A > B/A + 0.25$ (prolate off-plane), as spherical
all the bulges with both $B/A > 0.85$ and $B/A - 0.25 < C/A
< B/A + 0.25$, and as triaxial all the remaining bulges.
Spherical bulges will be treated as oblate spheroids in the analysis below.

\begin{figure}[t!]
\centering
\resizebox{\hsize}{!}{\includegraphics{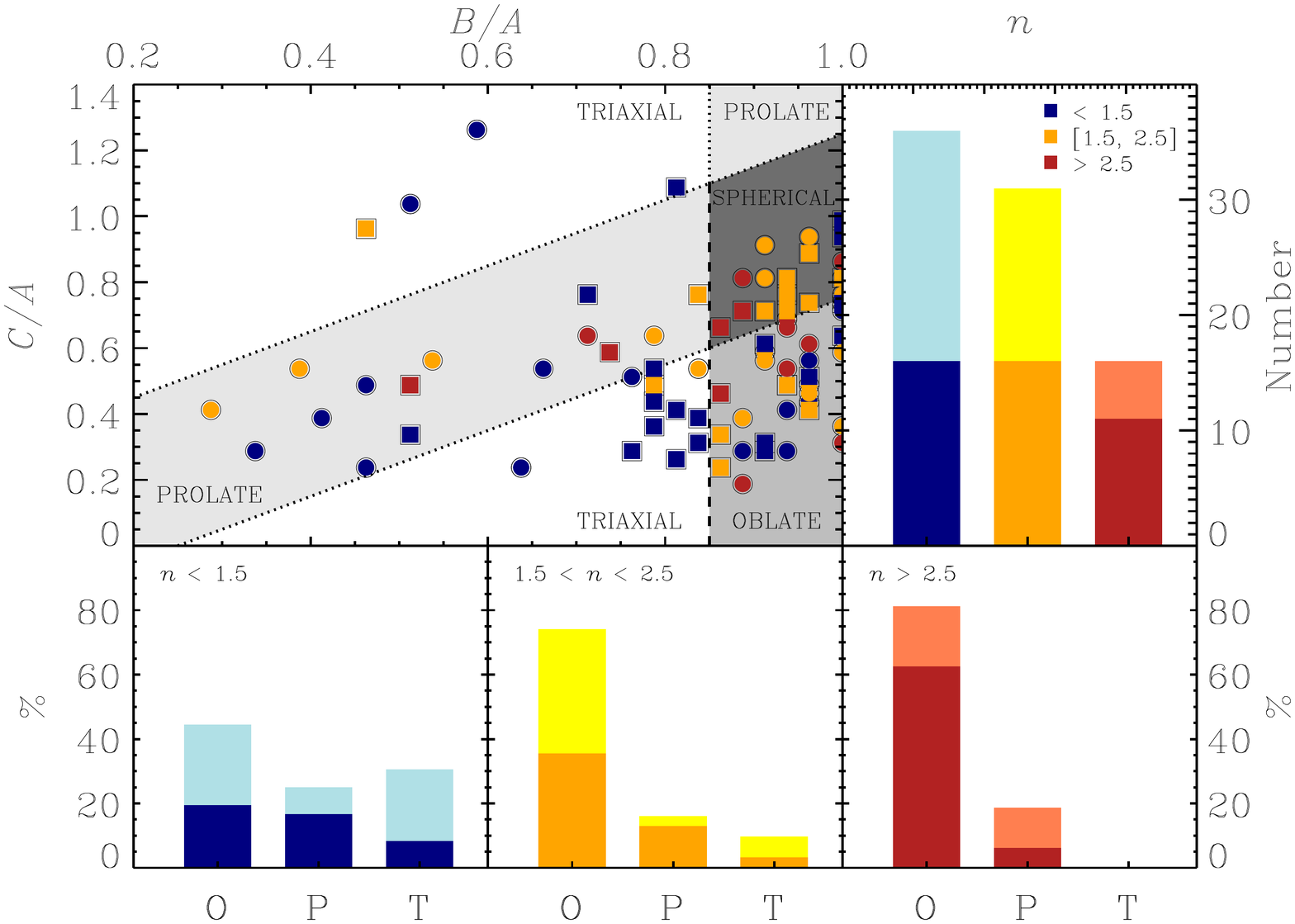}}
\caption{\emph{Top left panel:} Intrinsic axial ratios $B/A$ and $C/A$
of our CALIFA bulges as a function of their S\'ersic
index ($n < 1.5$: blue symbols; $1.5 < n < 2.5$: yellow symbols; $n
> 2.5$: red symbols). Circles and squares correspond to unbarred and
barred galaxies, respectively.  The white, light grey, grey, and dark grey regions
mark the regimes of triaxial, prolate, oblate, and spherical bulges,
respectively.  \emph{Top right panel:} Distribution of the S\'ersic
index of our CALIFA bulges ($n < 1.5$: blue histogram; $1.5 < n <
2.5$: yellow histogram; $n > 2.5$: red histogram). Dark and light
colors correspond to unbarred and barred galaxies,
respectively.  \emph{Bottom panels:} Distribution of the
intrinsic shape of our CALIFA bulges (O: oblate; P: prolate; T:
triaxial) as a function of their S\'ersic index ($n < 1.5$: blue
histograms; $1.5 < n < 2.5$: yellow histograms; $n > 2.5$: red
histograms). Dark and light colors correspond to unbarred and barred
galaxies, respectively.}
\label{fig:CALIFA_n}
\end{figure}

It is worth noting that 4 of our CALIFA bulges 
are oblate spheroids off-plane, while there are none prolate spheroids off-plane.
Such rare central structures swelling out the disk plane
have been recently studied by \citet{Corsini2012}, who found a slightly triaxial 
polar bulge with axial ratios $B/A = 0.95$ and $C/A = 1.60$ in \object{NGC~4698}.
We inspected the probability distribution of our 4 bulges and found 
that they presented a great scatter compatible also with being triaxial, as expected. 
Therefore, due to the peculiarity and the great uncertainty in the properties 
and formation mechanisms of polar bulges,
they should be considered as a particular kind of bulges 
and not include them in the main groups described in the forthcoming analysis.
Thus, the final sample of our CALIFA bulges comprises 79 objects 
(41 in unbarred galaxies and 38 in barred galaxies).

We distinguished all different bulges intrinsic shapes (oblate, prolate, or triaxial)
in the new $(B/A, C/A)$ diagram
according to the properties of their host galaxies.
As a general behaviour, we found that most of our CALIFA bulges tend
to be oblate ($66\%$), with a smaller fraction of prolate ($19\%$) or
triaxial bulges ($15\%$). The majority of triaxial bulges are in
barred galaxies ($75\%$). 
The $B/A$ and $C/A$ distribution peaks at $\langle B/A \rangle = 0.85$ and 
$\langle C/A \rangle = 0.55$, respectively.

\begin{figure}[t!]
\centering
\resizebox{\hsize}{!}{\includegraphics{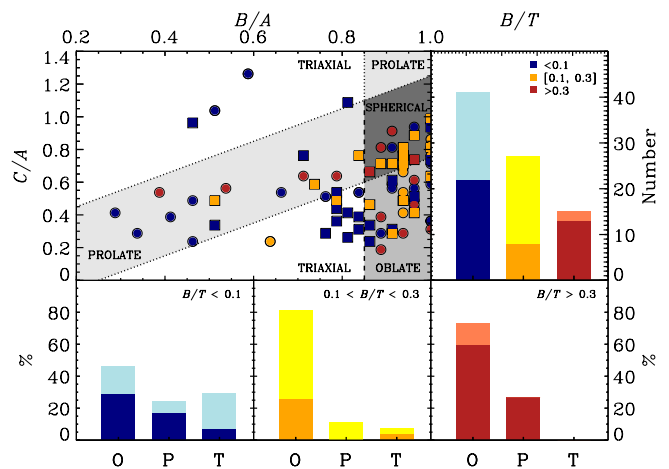}}
\caption{As in Fig.~\ref{fig:CALIFA_n}, but for the bulge-to-total
luminosity ratio. Our CALIFA bulges are divided in the following
bins: $B/T > 0.1$ (blue), $0.1 < B/T < 0.3$ (yellow), and $B/T >
0.3$ (red).}
\label{fig:CALIFA_BT}
\end{figure}

We divided our CALIFA bulges according to their S\'ersic index in the
bins $n \le 1.5$, $1.5 < n \le 2.5$, and $n>2.5$
(Fig.~\ref{fig:CALIFA_n}). 
The vast majority of our bulges ($80\%$) is
characterized by a small S\'ersic index ($n \le 2.5$). A substantial
fraction of bulges with $n>2.5$ ($69\%$) is observed in unbarred
galaxies. The bulges with $n \le 1.5$ have a variety of intrinsic
shapes, with comparable fractions of triaxial ($30\%$), oblate
($49\%$), and prolate bulges ($21\%$). By contrast, most of the bulges
with $1.5<n \le 2.5$ ($77\%$) and $n>2.5$ ($81\%$) are
oblate. Finally, we noticed that almost all the triaxial bulges show
very small values of S\'ersic index ($n<1.5$). The same trends were
seen by dividing our CALIFA bulges in the bins $B/T \le 0.1$, $0.1<B/T
\le 0.3$, and $B/T \le 0.3$, according to the bulge-to-total
luminosity ratio of their host galaxy (Fig.~\ref{fig:CALIFA_BT}). Most
of the larger bulges are oblate (74\%),
while the smaller ones show a variety of intrinsic shapes. We also
pointed out that the bulges with small values of intrinsic flattening
$C/A$ have systematically small values 
of $n$ and $B/T$.

\begin{figure}[t!]
\centering
\resizebox{\hsize}{!}{\includegraphics{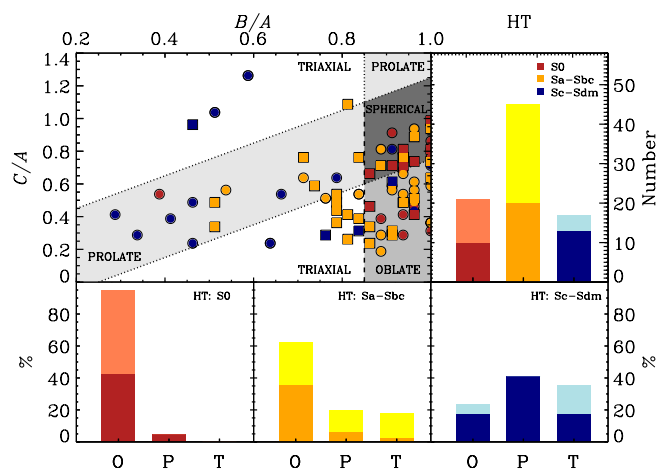}}
\caption{As in Fig.~\ref{fig:CALIFA_n}, but for the Hubble type.  
Our CALIFA bulges are divided in the following bins: S0 (red),
Sa--Sbc (yellow), Sc--Sdm (blue).}
\label{fig:CALIFA_HT}
\end{figure}

We also analysed the bulge intrinsic shape to highlight possible
correlations with the morphology of the host galaxy
(Fig. \ref{fig:CALIFA_HT}). We separated our CALIFA bulges into three
bins by taking into account the bulges in S0 galaxies, bulges in Sa,
Sab, Sb, and Sbc galaxies, and bulges in Sc, Scd, Sd, and Sm
galaxies. Most of the bulges belong to galaxies in the 
Sa--Sbc bin (56\%). Almost all the bulges in S0 galaxies
($95\%$) are oblate, with a different degree of intrinsic flattening
$C/A$. We did not find any triaxial bulge among the S0 galaxies. On
the contrary, the bulges of spiral galaxies present a variety of
intrinsic shapes, with oblate bulges ($62\%$) dominating the Sa--Sbc
bin. Moreover, we noticed that bulges with small values of $C/A$ are
more frequently observed in late-type spirals.

\begin{figure}[t!]
\centering
\resizebox{\hsize}{!}{\includegraphics{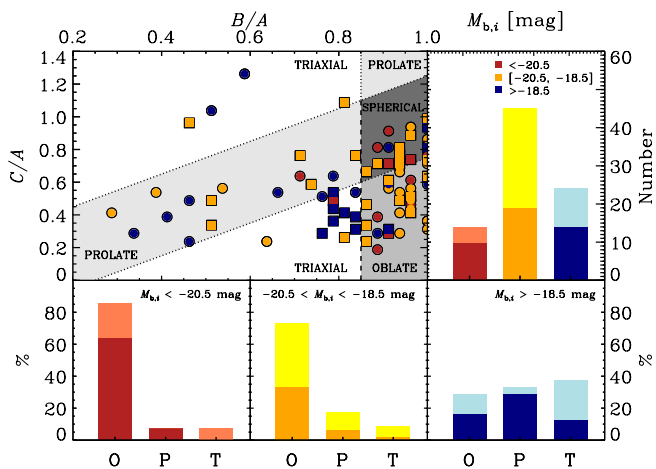}}
\caption{As in Fig.~\ref{fig:CALIFA_n}, but for the absolute magnitude
of the bulge. Our CALIFA bulges are divided in the following bins:
$M_{{\rm b}, \, i} < -20.5$ mag (red), $-20.5 < M_{{\rm b}, \, i} <
-18.5$ mag (yellow), and $M_{{\rm b}, \, i} > -18.5$ mag (blue).}
\label{fig:CALIFA_Mb}
\end{figure}

\begin{figure}[t!]
\centering
\resizebox{\hsize}{!}{\includegraphics{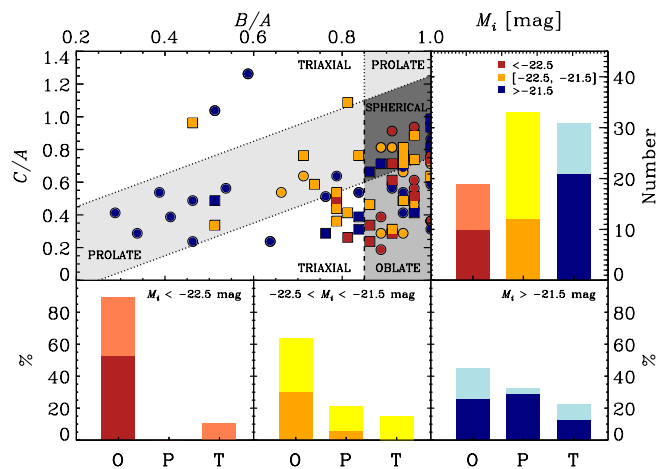}}
\caption{As in Fig. \ref{fig:CALIFA_n}, but for the absolute magnitude
of the galaxy. Our CALIFA bulges are divided in the following bins:
$M_{i} < -22.5$ mag (red), $-22.5 < M_{i} < -21.5$ mag (yellow), and
$M_{i} > -21.5$ mag (blue).}
\label{fig:CALIFA_Mi}
\end{figure}

Finally, we studied the bulge intrinsic shape as a function of
$i$-band absolute magnitude of the bulge (Fig.~\ref{fig:CALIFA_Mb})
and of the host galaxy (Fig. \ref{fig:CALIFA_Mi}). 
Almost all the most massive bulges ($M_{{\rm b}, \, i} < -20.5$ mag) 
are oblate ($86\%$), whereas the less massive one 
($M_{{\rm b}, \, i} > -18.5$ mag) are more heterogeneous with a similar 
fraction of triaxial ($41\%$), oblate ($27\%$), and prolate systems ($32\%$). 
We obtained the same results when the total galaxy absolute magnitude 
was examined.


\section{Discussion \label{sec:discussion}}

The statistical analysis presented in this work allowed us
to individually constrain the intrinsic shape of a sample of bulges in relation to
their observed properties. We projected the ($B/A$, $C/A$) values
in order to compare the bulge shape distribution 
with previous results (see Fig.~\ref{fig:shape_tot}).
We found that the mean axial ratio
of our CALIFA bulges is $\langle B/A \rangle = 0.85$ and 
$\langle C/A \rangle = 0.55$, respectively. This result is in agreement with 
previous analyses by \citet{Bertola1991a}, \citet{Fathi2003}, and \citet{Mendez2008}.

Since the cumulative projected distribution 
 mixes all different shapes (oblate, prolate, and triaxial),
we preferred to distinguish the properties of our CALIFA 
bulges in the ($B/A, C/A$) diagram.
Indeed, the actual position of bulges in the ($B/A, C/A$) diagram
is a powerful tool for disentangling bulge types.
We found that some of our CALIFA bulges (6\%) are very
flattened oblate systems ($B/A >0.85$ and $C/A < 0.3$),
which are possible candidate to be disk-like bulges.
Moreover, since barred galaxies are found to host the majority of
triaxial bulges, they could be interpreted as the signature of boxy/peanut
structures. Indeed, the secular evolution of the bar via
buckling or resonants effects is known to result in thick triaxial
components. Even the inclusion of the bar in the 
photometric decomposition can not avoid a mild 
contamination from boxy/peanut structures. 
Thus, it is not surprising that barred galaxies
show a large fraction of triaxial bulges.
It is worth noting that in discussing the shape of
the sample bulges obtained from the ($B/A, C/A$) diagram, it was
considered the statistical meaning of the intrinsic axial ratios
we derived and the empiric definition we adopted for the oblate,
spherical, prolate, and triaxial bulges. The $1\sigma$ contour
level of the distribution of the intrinsic axial ratios of a bulge
can tightly or loosely circle the most probable values of its
shape (see Fig.~\ref{fig:simulations_shape}, for an example). 
Therefore, a certain degree of
triaxiality is allowed also for the bulges we classified as
oblate or prolate. On the other hand, the definition of the
boundaries of the regions marking the different bulge shapes in
the ($B/A, C/A$) diagram might be very conservative.

These results are consistent with a major role of a certain mechanism 
in the buildup of the most massive bulges (usually identified by higher $n$, 
higher $B/T$, earlier types, more massive systems) 
which has not significantly contributed to those forming the less 
massive ones (usually identified by lower $n$, lower $B/T$, 
later types and in less massive galaxies).
Some evolutionary mechanisms 
may have taken place in all mass ranges similarly 
(such as internal secular evolution or cluster-related processes), 
but it is obvious from these results that there are some specific processes 
that have contributed much more in the most massive bulges 
to make them more homogeneous in shape (i.e., all oblate)
than in the bulges with lower masses. 
These processes must impose over others and have occurred 
more frequently in massive systems than in less massive ones,
as well as they have also contributed to increase $n$ and $B/T$ in the galaxy at the same time, 
i.e., they must transform the system towards an earlier type. 

In the less massive bulges, the interplay of different
evolutionary mechanisms can explain the wide variety of
shapes, as they can be more or less relevant in a galaxy depending
on its evolutionary history and environment. However, the presence
of the bar seems to drive the evolution of
low-mass triaxial bulges. Indeed, triaxial bulges are
mostly hosted in barred galaxies with low values of $B/T$,
$M_{{\rm b}, \, i}$, and $n$. These bulges could be 
contaminated by the residual light of the
low-inclined counterparts of boxy/peanut structures. The lack of
triaxial bulges in lenticular barred galaxies could be explained by
the larger mass of their bulges: their deep
potential well seems to reshape the central region into a
more axisymmetric structure, where the bar has a marginal
role in perturbing the bulge. Thus, the bulge mass could play a
role also in driving the evolution of bulges in barred galaxies.

Many studies report observational evidence on a major role of both
major and minor merging and dissipative collapse in the buildup of the most massive galaxies 
\citep[e.g.,][]{Rudick2009, ElicheMoral2010, Thomas2010, Kaviraj2011, Bernardi2011, 
Mendez2012, Barway2013, Prieto2013, Morelli2016, Leja2015, Prieto2015}.
Numerical simulations have shown that gas-poor major and minor mergers 
tend to introduce some triaxiality in bulges that originally were spheroidal 
\citep{Cox2006, Tapia2014}. However, the bulges of dry 
minor-merger remnants also exhibit higher rotational support at their centers, 
even though the global rotational support of the galaxy decreases,
making the bulge more oblate \citep{Tapia2014}. 
This happens because part of the orbital angular momentum of the encounter 
is transferred to the inner regions \citep[see][]{ElicheMoral2006, ElicheMoral2011},
contributing to the flattening of the material at the galaxy center. 
High gas amounts in the progenitors only contribute to make the remnant 
more axisymmetric \citep{Jesseit2007},
so the trend of dry mergers to make remnant bulges more 
oblate can be extrapolated to wet ones. 
Therefore, our results would be consistent with a higher relevance of merging 
in the formation and evolution of the most massive bulges.

\section{Conclusions \label{sec:conclusions}}

We derived the intrinsic shape of 83 bulges of a sample of nearby
galaxies from CALIFA DR3. To this aim we applied the statistical
method by \citet{Mendez2008} to the structural parameters obtained by
\citet{Mendez2017} with a two-dimensional photometric decomposition of
the SDSS $i$-band images of the sample galaxies.

We made use of a set of simulated galaxies resulting from merger
experiments, that closely resembling lenticular galaxies, to test the
reliability of the method by \citet{Mendez2008}. For each simulated
lenticular remnant, we created a set of mock SDSS $i$-band images at different
galaxy inclinations to mimic the observing setup of  
SDSS images of CALIFA DR3 galaxies.
We performed a two-dimensional photometric decomposition
of all the mock images applying the same procedure as for real galaxies,
in order to retrieve the geometrical parameters
of bulge and disk which we used to recover the bulge intrinsic
shape. The probability distributions of the axial ratios $B/A$ and
$C/A$ obtained for different inclinations for the same simulated
lenticular remnant overlap at $1\sigma$ level. We concluded that the adopted method
allows us to successfully constrain the bulge intrinsic shape when the
galaxy inclination is $25^{\circ} < \theta < 65^{\circ}$. We also
realized that a very accurate photometric decomposition is mandatory
to retrieve the bulge intrinsic shape and that a galaxy inclination of
$\theta = 60^{\circ}$ returns the tightest constraints on the
intrinsic axial ratios $B/A$ and $C/A$ of the bulge.

We divided our CALIFA bulges according to their intrinsic axial ratios
$B/A$ and $C/A$ into oblate (in-plane or off-plane), 
prolate (in-plane or off-plane), and triaxial.
We looked for possible correlations
between the intrinsic shape of our bulges and some of the basic
properties of their host galaxies (i.e., S\'ersic index of the bulge
$n$, bulge-to-total luminosity ratio $B/T$, Hubble type HT, $i$-band
absolute magnitude of the bulge $M_{{\rm b}, \, i}$, and $i$-band
absolute magnitude of the galaxy $M_i$) as derived by
\citealt{Mendez2017}. Our analysis pointed out that bulges with a
small value of $n$ or $B/T$ could be equally axisymmetric or triaxial
ellipsoids, while most of the bulges with large values of $n$ or $B/T$ are mostly
oblate spheroids. Moreover, less massive bulges and bulges in
late-type galaxies presented heterogeneous intrinsic shapes, 
while more massive bulges and bulges in lenticular galaxies are mostly oblate. 
Finally, we found the majority of triaxial bulges in barred galaxies.

We concluded that merging events and dissipative collapse
could be responsible of driving the formation and evolution of our
most massive bulges, although other physical mechanisms, i.e., 
the internal secular evolution caused by the presence of the bar, may be
acting at the same time.  The
coexistence of different pathways is more clear in less
massive bulges, where the bar seems to reshape low-mass
triaxial bulges.
In this context, the role of simulations result crucial in unveiling 
various evolution pathways in nearby galaxies.
Nevertheless, very few numerical studies have focused 
on the bulge evolution and in particular on the intrinsic shape.
Thus, our results imposed further limitations on forthcoming 
numerical simulations and may help to disentangle different formation scenarios.

\begin{acknowledgements}

We would like to thank the anonymous referee for the suggestions 
that helped us to improve the way we presented our results.

This work is supported by Padua University through grants 60A02-4434/15,
DOR1699945, and DOR1715817. E.M.C. and L.M. acknowledge financial
support from Padua University through grants BIRD164402/16 an CPS0204,
respectively. J.M.A. want to thank the support of this
work by the Spanish Ministerio de Economia y Competitividad (MINECO) 
under the grant AYA2013-43188-P.
L.C. thanks the IAC for hospitality while this
paper was in progress. We would also like to acknowledge
I.~Chilingarian, P.~Di Matteo, F.~Combes, A.-L.~Melchior, and
B.~Semelin for creating the GalMer database, and the HORIZON project
for supporting it (\url{http://www.projet-horizon.fr/\-rubrique3.html}).
This research made use of Sloan Digital Sky Survey (SDSS) data
(\url{http://www.sdss.org}) and NASA/IPAC Extragalactic Database (NED)
(\url{http://ned.ipac.caltech.edu/}).

\end{acknowledgements}


\newpage
\newpage

\begin{appendix}

\section{\label{appendix:A}}

\begin{table}[h!]
\centering
\caption{Structural parameters of the simulated lenticular remnants resulting
from numerical experiments of giant-giant galaxy mergers.}
\begin{adjustbox}{max width=\linewidth}
\centering
\begin{tabular}{cccccccc}
\hline
\hline
\noalign{\smallskip}
Galaxy & $\theta$ & $r_{\rm e}$ & $n$ & $h$ & $q_{\rm b}$ & $q_{\rm d}$ & $|\delta|$\\
    & [$^\circ$] & [arcsec] & & [arcsec]	&  &  & [$^\circ$]\\
(1) & (2) & (3)	& (4) & (5) & (6) & (7) & (8)  \\
\noalign{\smallskip}
\hline
\noalign{\smallskip}
\multirow{5}{*}{gE0gSbo5}& 0 & 0.8 & 3.8 & 9.0 & 0.95 & 0.98 & 90\\
&180 & 0.8 & 3.8 & 9.0 & 1.00 & 0.98      & 32\\
& 30 & 0.9 & 4.9 & 9.1 & 1.00 & 0.97      & 19\\
& 45 & 0.9 & 3.9 & 9.4 & 0.86 & 0.78$^{*}$ & 4\\
& 60 & 1.0 & 4.1 & 9.7 & 0.78 & 0.67$^{*}$ & 1\\
\noalign{\smallskip}
\hline
\noalign{\smallskip}
\multirow{5}{*}{gE0gSdo5}& 0 & 1.0 & 1.1 & 8.7 & 0.80 & 0.99 & 6\\
&180 & 1.0 & 1.1 & 8.7 & 0.80 & 0.99       & 6\\
& 30 & 0.9 & 1.2 & 8.7 & 0.90 & 0.90$^{*}$ & 23\\
& 45 & 0.9 & 1.3 & 8.9 & 0.81 & 0.83$^{*}$ & 18\\
& 60 & 0.8 & 1.5 & 9.1 & 0.64 & 0.70$^{*}$ & 7\\
\noalign{\smallskip}
\hline
\noalign{\smallskip}
\multirow{5}{*}{gSbgSbo9}& 0 & 1.3 & 3.5 &10.9 & 0.72 & 0.97 & 90\\
&180 & 1.3 & 3.5 & 10.9 & 0.72 & 0.97 & 90\\
& 30 & 1.2 & 3.8 & 11.0 & 0.78 & 0.87 & 63\\
& 45 & 1.1 & 4.2 & 11.3 & 0.81 & 0.74 & 42\\
& 60 & 1.1 & 4.9 & 11.6 & 0.73 & 0.57 & 18\\
\noalign{\smallskip}
\hline
\noalign{\smallskip}
\multirow{5}{*}{gSbgSdo5}& 0 & 2.3 & 3.6 & 18.0 & 0.83 & 0.99 & 58\\
&180 & 2.4 & 4.0 & 18.1 & 0.84 & 0.99 & 59\\
& 30 & 2.3 & 3.8 & 18.0 & 0.82 & 0.92 & 29$^{*}$\\
& 45 & 2.2 & 3.6 & 17.4 & 0.75 & 0.82 & 21$^{*}$\\
& 60 & 2.1 & 3.5 & 17.4 & 0.64 & 0.67 & 12$^{*}$\\
\noalign{\smallskip}
\hline
\end{tabular}
\end{adjustbox}
\tablefoot{(1) Identifier of the simulated lenticular remnant. (2) Galaxy
inclination. (3) Effective radius of the bulge. (4)
S\'ersic index of the bulge.  (5) Scale length of the disk. (6), (7)
Apparent axial ratio of the bulge and disk, respectively.  (8)
Difference of the position angles of bulge and disk. Nominal values
are marked with $^{*}$. 
}
\label{tab:photometry_1}
\end{table}

\begin{table}[h!]
\caption{As in Table~\ref{tab:photometry_1}, but simulated lenticular remnants
resulting from numerical experiments of giant-dwarf galaxy mergers.}
\centering
\begin{adjustbox}{max width=\linewidth}
\begin{tabular}{cccccccc}
\hline
\hline
\noalign{\smallskip}
Galaxy & $\theta$ & $r_{\rm e}$ & $n$ & $h$ & $q_{\rm b}$ & $q_{\rm d}$ & $|\delta|$\\
    & [$^\circ$] & [arcsec] & & [arcsec]	&  &  & [$^\circ$]\\
(1) & (2) & (3)	& (4) & (5) & (6) & (7) & (8)  \\

\noalign{\smallskip}
\hline
\noalign{\smallskip}
\multirow{5}{*}{gS0dS0o99}& 0 & 7.0 & 1.1 & 18.9 & 0.60 & 0.97 & 59\\
&180 & 7.0 & 1.1 & 19.2 & 0.60 & 0.96 & 82\\
& 30 & 6.8 & 1.1 & 18.9 & 0.59 & 0.89 & 33\\
& 45 & 6.6 & 1.1 & 18.8 & 0.57 & 0.76 & 18\\
& 60 & 6.3 & 1.1 & 18.5 & 0.53 & 0.59 & 11\\
\noalign{\smallskip}
\hline
\noalign{\smallskip}
\multirow{5}{*}{gS0dSao103}& 0 & 7.7 & 1.5 & 22.1 & 0.54 & 0.97 & 20\\
&180 & 7.8 & 1.5 & 22.1 & 0.54 & 0.98 & 20\\
& 30 & 7.4 & 1.5 & 21.6 & 0.56 & 0.93 & 46\\
& 45 & 6.9 & 1.6 & 21.5 & 0.60 & 0.80 & 47\\
& 60 & 6.2 & 1.6 & 21.1 & 0.63 & 0.63 & 37\\
\noalign{\smallskip}
\hline
\noalign{\smallskip}
\multirow{5}{*}{gS0dSdo106}& 0 & 7.7 & 1.1 & 24.1 & 0.46 & 0.93 & 83\\
&180 & 7.7 & 1.0 & 24.1 & 0.46 & 0.93 & 83\\
& 30 & 7.4 & 1.1 & 22.2 & 0.46 & 0.92 & 36\\
& 45 & 6.8 & 1.0 & 21.2 & 0.45 & 0.77 & 20\\
& 60 & 8.3 & 1.3 & 30.1 & 0.43 & 0.59 & 9\\
\noalign{\smallskip}
\hline
\noalign{\smallskip}
\multirow{5}{*}{gS0dSdo100}& 0 & 7.9 & 1.6 & 26.3 & 0.53 & 0.93 & 73\\
&180 & 7.9 & 1.6 & 26.3 & 0.53 & 0.93 & 74\\
& 30 & 7.9 & 1.6 & 24.9 & 0.50 & 0.91 & 13\\
& 45 & 8.0 & 1.7 & 25.1 & 0.47 & 0.76 & 4\\
& 60 & 8.3 & 1.7 & 25.7 & 0.44 & 0.56 & 1\\
\noalign{\smallskip}
\hline
\end{tabular}
\end{adjustbox}
\label{tab:photometry_2}
\end{table}

\begin{table}[t!]
\caption{As in Table~\ref{tab:photometry_1}, but simulated lenticular remnants
resulting from numerical experiments of (giant S0)-(dwarf E0) galaxy
mergers with different orbital parameters.}
\centering
\begin{adjustbox}{max width=\linewidth}
\begin{tabular}{cccccccc}
\hline
\hline
\noalign{\smallskip}
Galaxy & $\theta$ & $r_{\rm e}$ & $n$ & $h$ & $q_{\rm b}$ & $q_{\rm d}$ & $|\delta|$\\
    & [$^\circ$] & [arcsec] & & [arcsec]	&  &  & [$^\circ$]\\
(1) & (2) & (3)	& (4) & (5) & (6) & (7) & (8)  \\
\noalign{\smallskip}
\hline
\noalign{\smallskip}
\multirow{5}{*}{gS0dE0o98}& 0 & 7.8 & 1.1 & 21.5 & 0.49 & 0.90 & 9\\
&180 & 7.9 & 1.1 & 21.6 & 0.49 & 0.91 & 13\\
& 30 & 7.4 & 1.1 & 21.7 & 0.52 & 0.82 & 30\\
& 45 & 6.8 & 1.2 & 21.9 & 0.54 & 0.70 & 26\\
& 60 & 6.4 & 1.2 & 22.4 & 0.54 & 0.53 & 19\\
\noalign{\smallskip}
\hline
\noalign{\smallskip}
\multirow{5}{*}{gS0dE0o99}& 0 & 7.8 & 1.1 & 20.8 & 0.49 & 0.99 & 16\\
&180 & 7.8 & 1.1 & 20.7 & 0.49 & 0.97 & 16\\
& 30 & 6.7 & 1.1 & 18.5 & 0.56 & 0.91 & 7\\
& 45 & 5.9 & 1.2 & 18.5 & 0.66 & 0.89 & 54\\
& 60 & 4.9 & 1.2 & 18.7 & 0.76 & 0.66 & 40\\
\noalign{\smallskip}
\hline
\noalign{\smallskip}
\multirow{5}{*}{gS0dE0o100}& 0 & 8.0 & 1.1 & 21.2 & 0.48 & 0.85 & 1\\
&180 & 8.0 & 1.1 & 21.1 & 0.47 & 0.87 & 1\\
& 30 & 7.3 & 1.2 & 20.9 & 0.53 & 0.87 & 38\\
& 45 & 6.6 & 1.2 & 22.1 & 0.59 & 0.74 & 44\\
& 60 & 5.8 & 1.2 & 22.9 & 0.65 & 0.56 & 31\\
\noalign{\smallskip}
\hline
\end{tabular}
\end{adjustbox}
\label{tab:photometry_3}
\end{table}

\mbox{}

\section{ \label{appendix:B}}

\begin{figure}[h!]
\centering
\resizebox{\hsize}{!}{\includegraphics{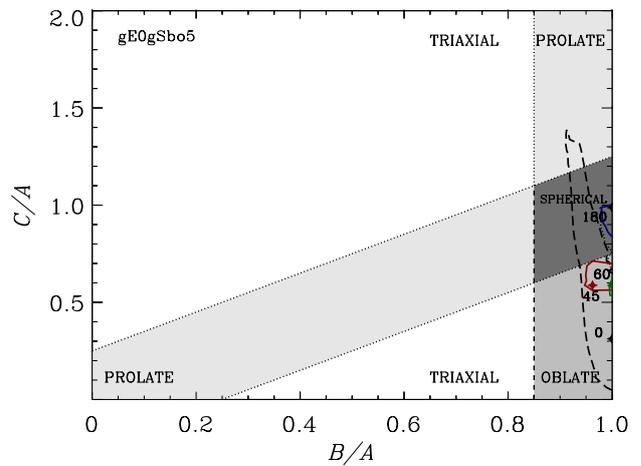}}
\caption{As in Fig.~\ref{fig:simulations_shape}, but for the remaining simulated lenticular remnants.
The most probable ($B/A, C/A$) values for the simulated lenticular remnant 
gE0gSbo5 at $\theta = 180^{\circ}$ and $\theta = 30^{\circ}$ practically overlap.}
\label{fig:simulations_shape_B}
\end{figure}

\begin{figure}[h!]
\centering
\addtocounter{figure}{-1}
\resizebox{\hsize}{!}{\includegraphics{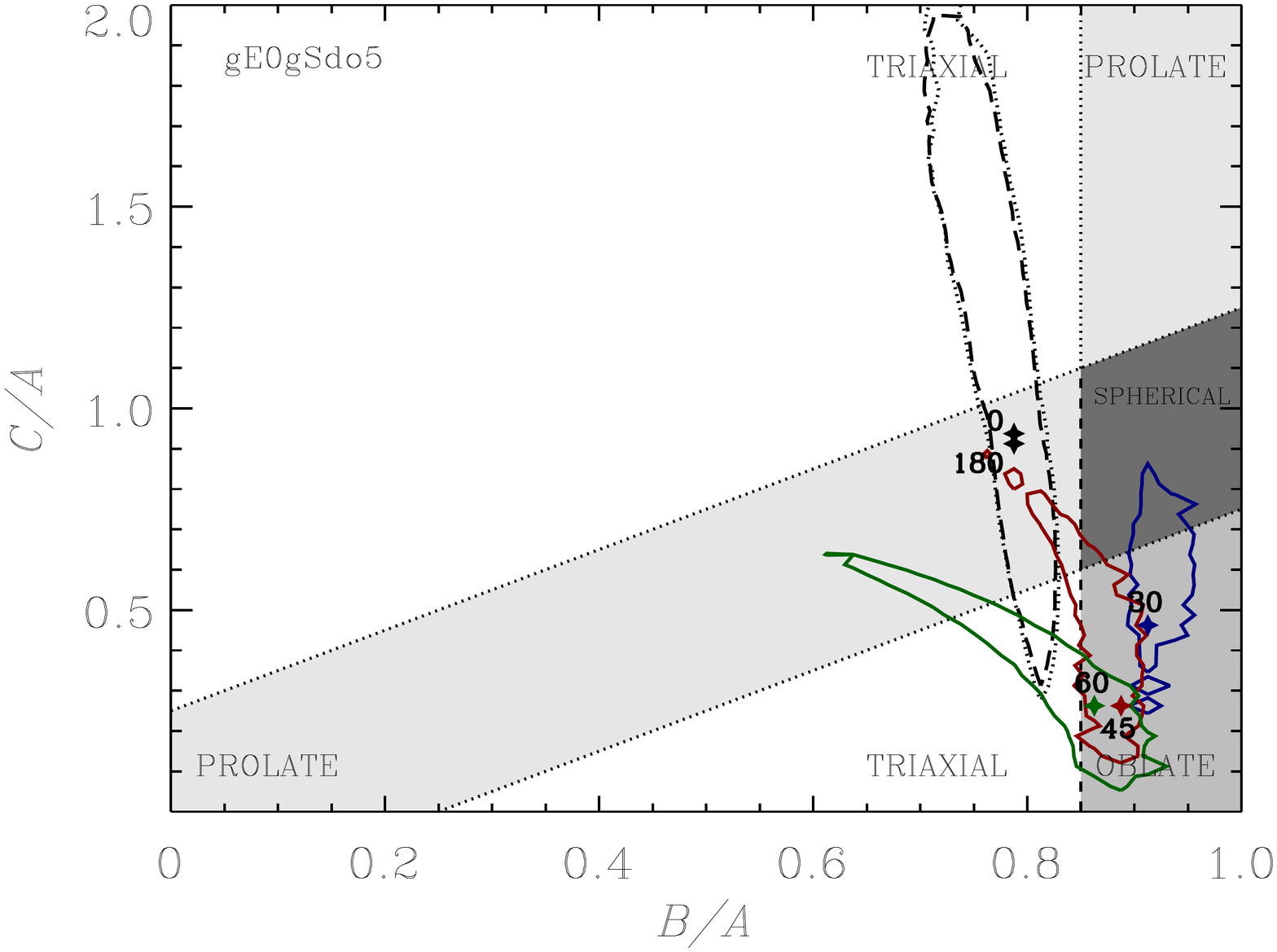}}
\resizebox{\hsize}{!}{\includegraphics{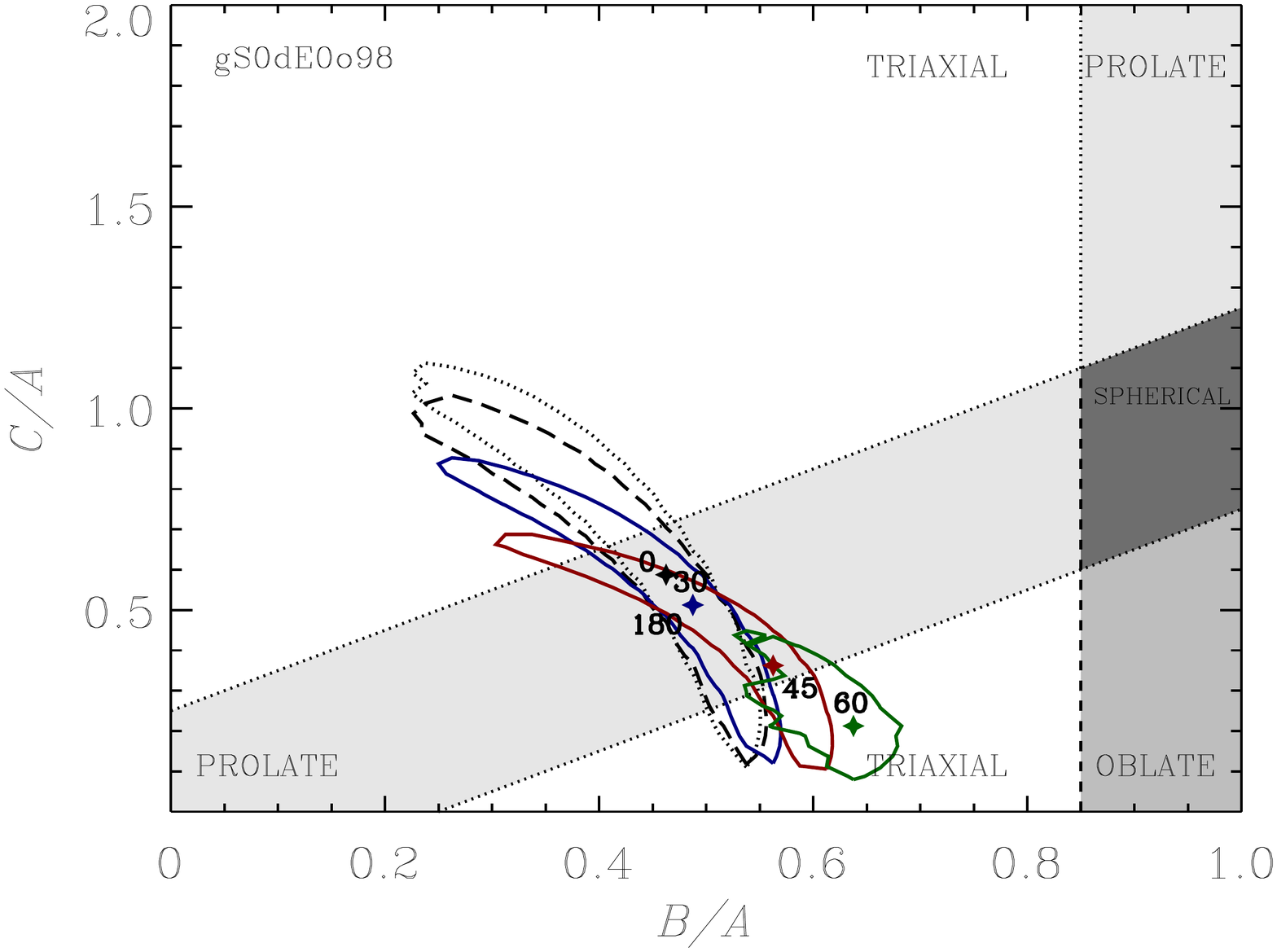}}
\resizebox{\hsize}{!}{\includegraphics{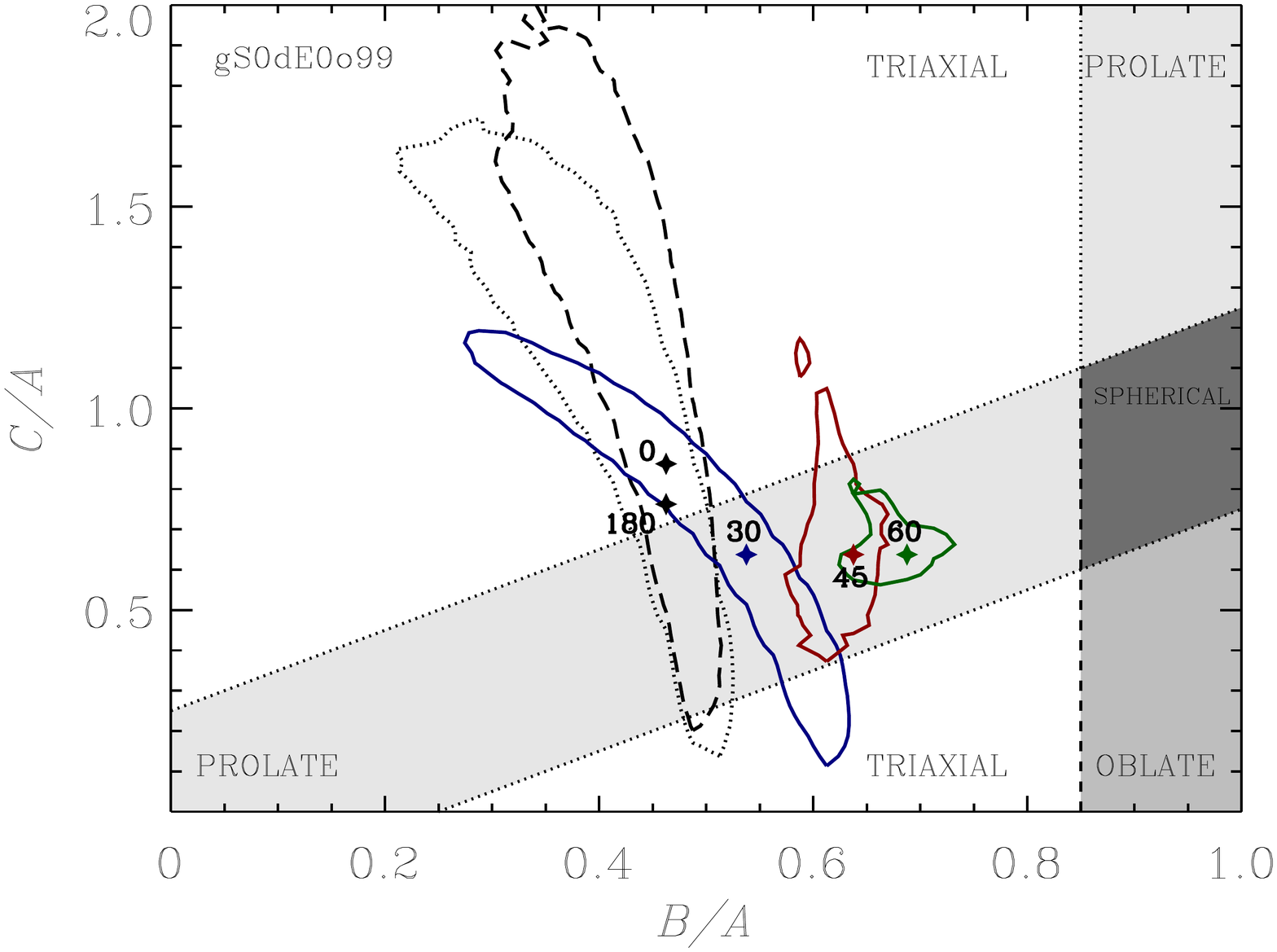}}
\caption{continued.}
\end{figure}

\begin{figure}[h!]
\centering
\addtocounter{figure}{-1}
\resizebox{\hsize}{!}{\includegraphics{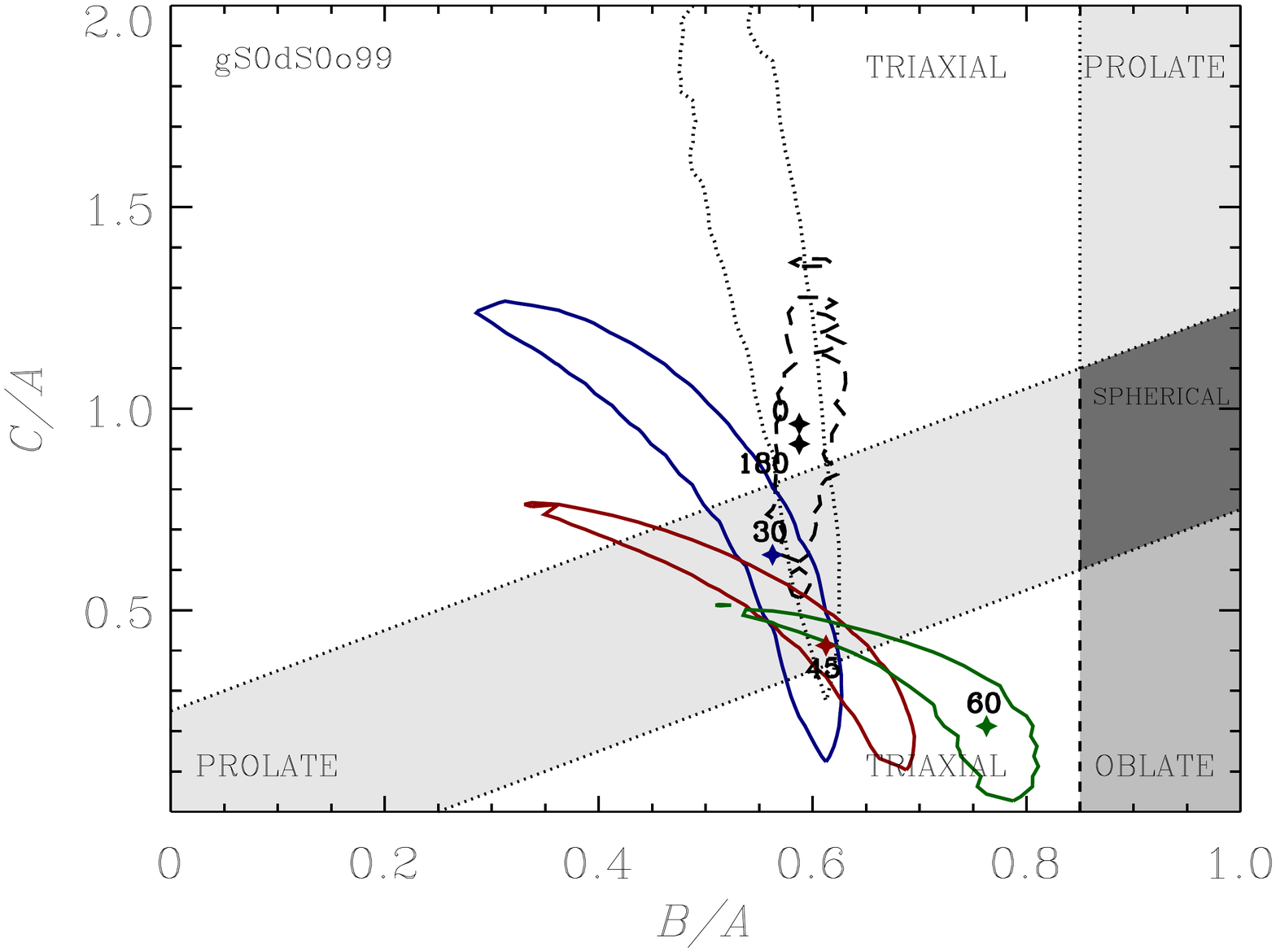}}
\resizebox{\hsize}{!}{\includegraphics{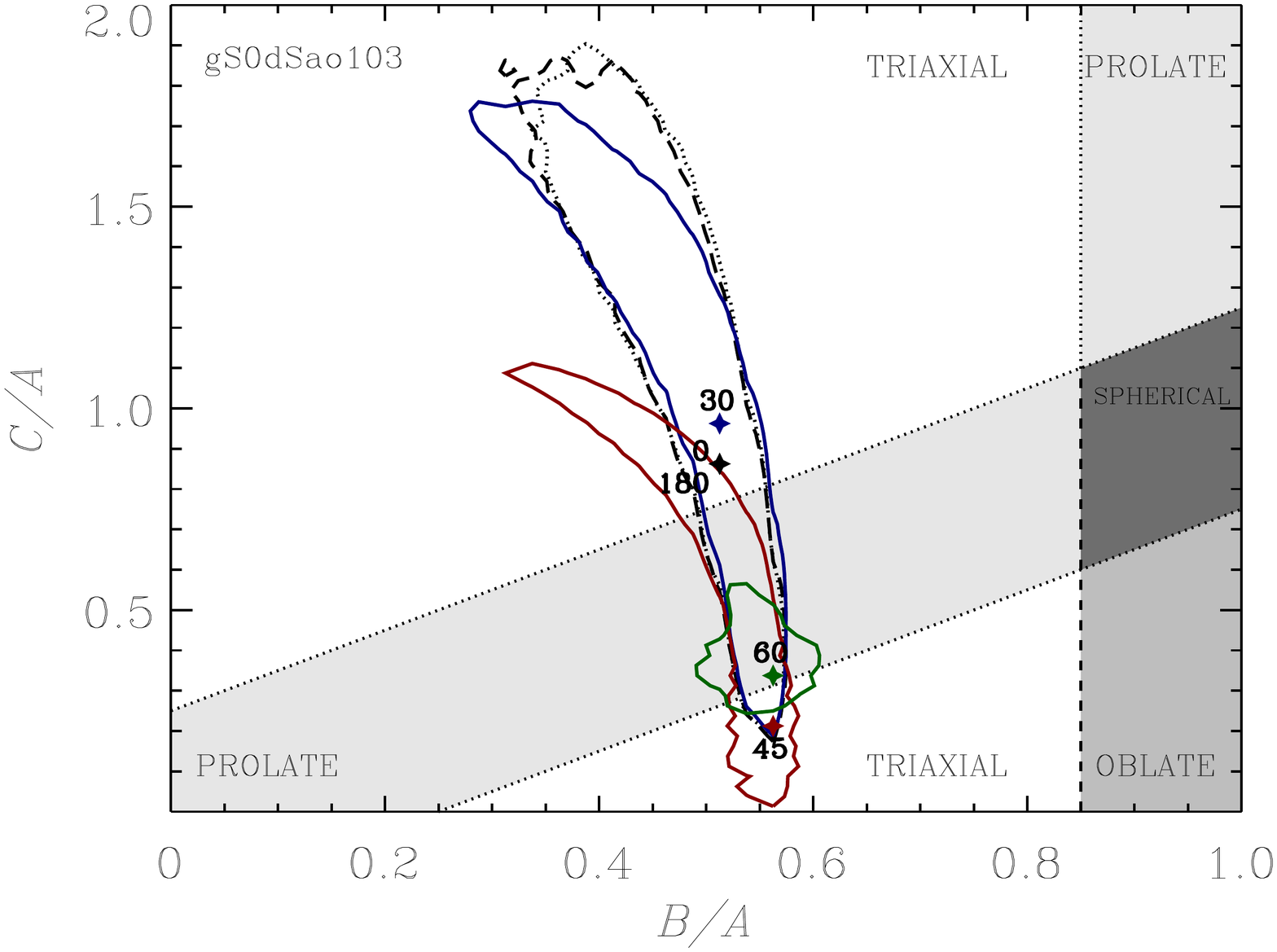}}
\resizebox{\hsize}{!}{\includegraphics{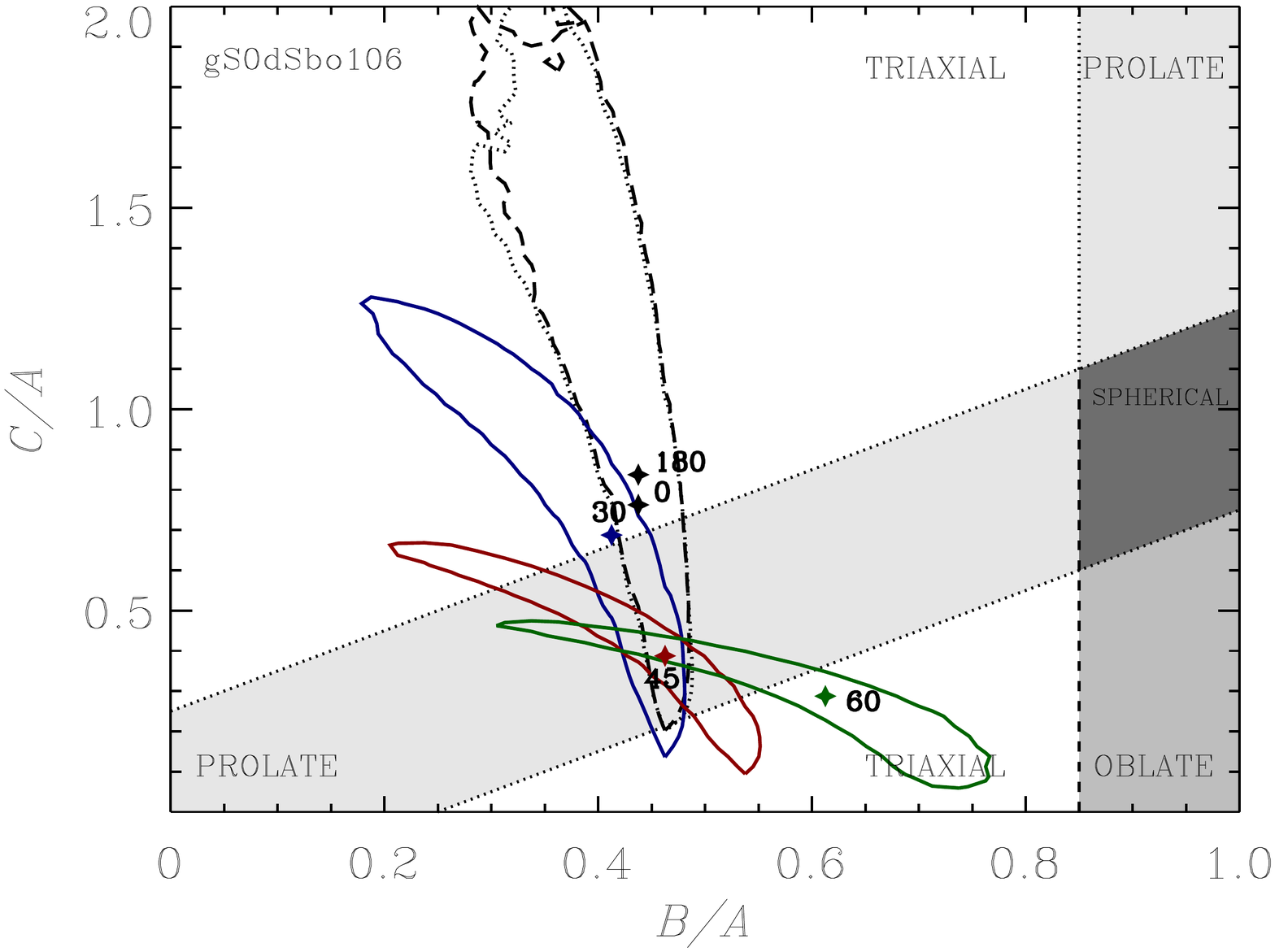}}
\caption{continued.}
\end{figure}

\begin{figure}[h!]
\centering
\addtocounter{figure}{-1}
\resizebox{\hsize}{!}{\includegraphics{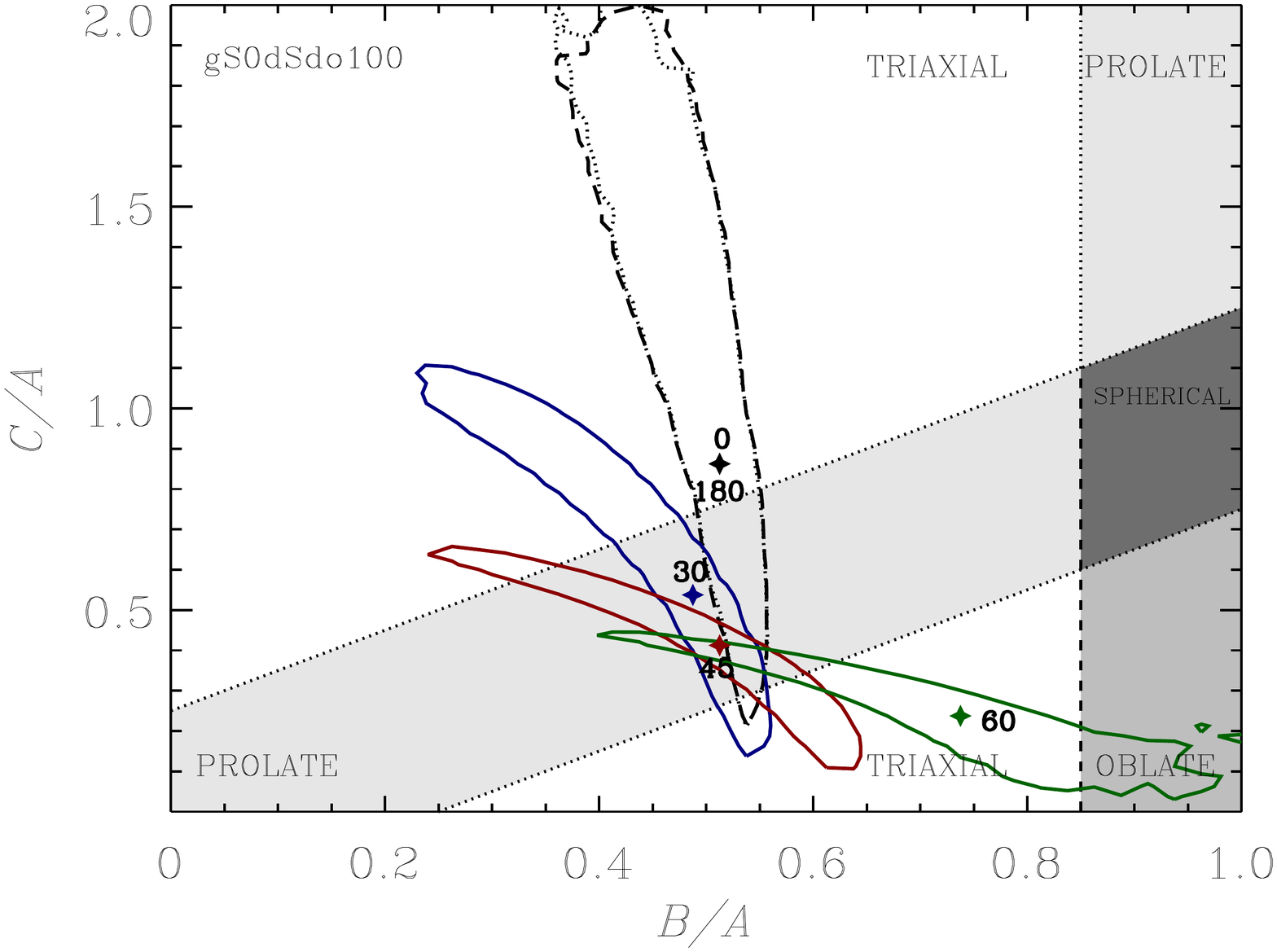}}
\resizebox{\hsize}{!}{\includegraphics{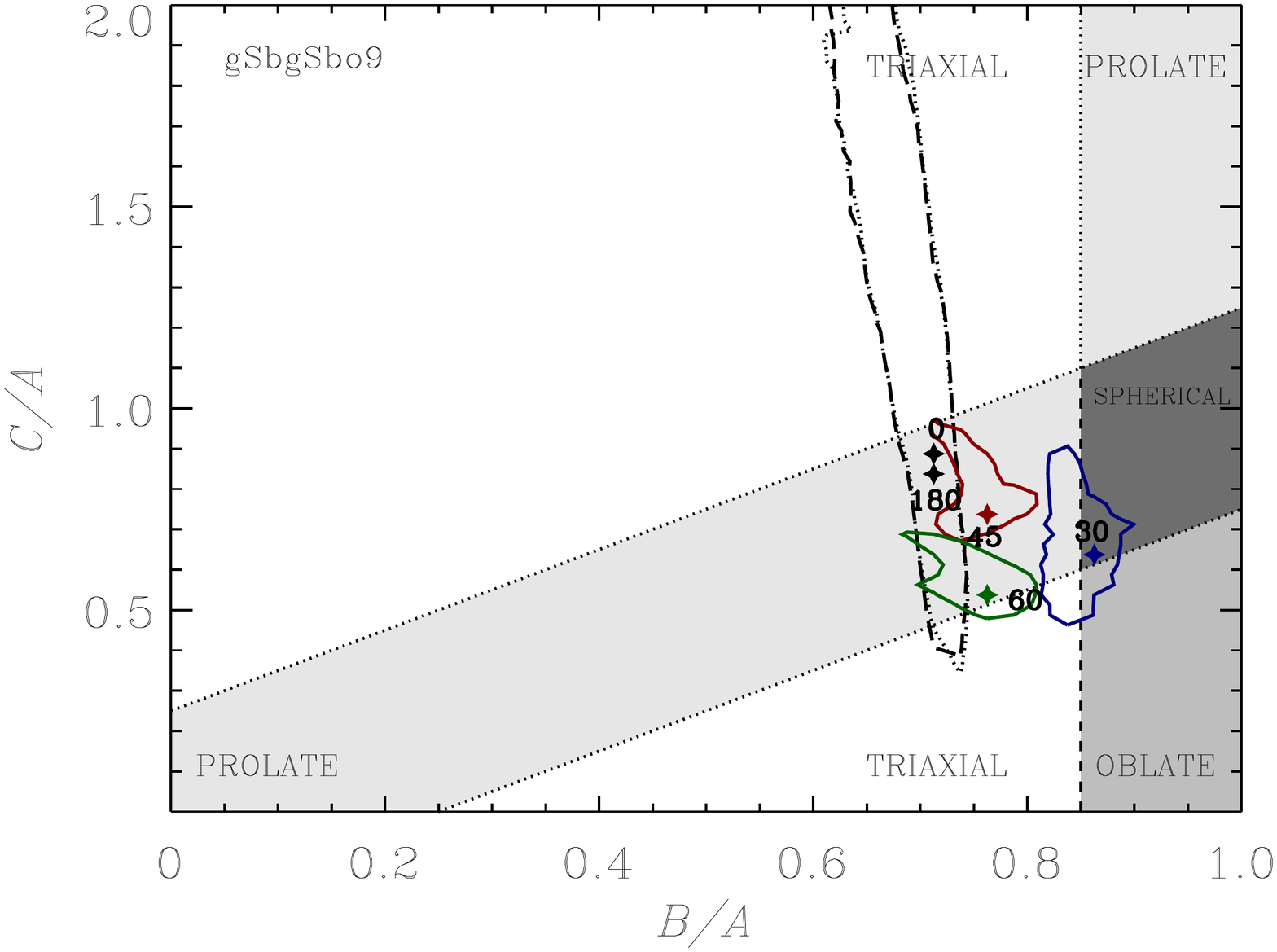}}
\resizebox{\hsize}{!}{\includegraphics{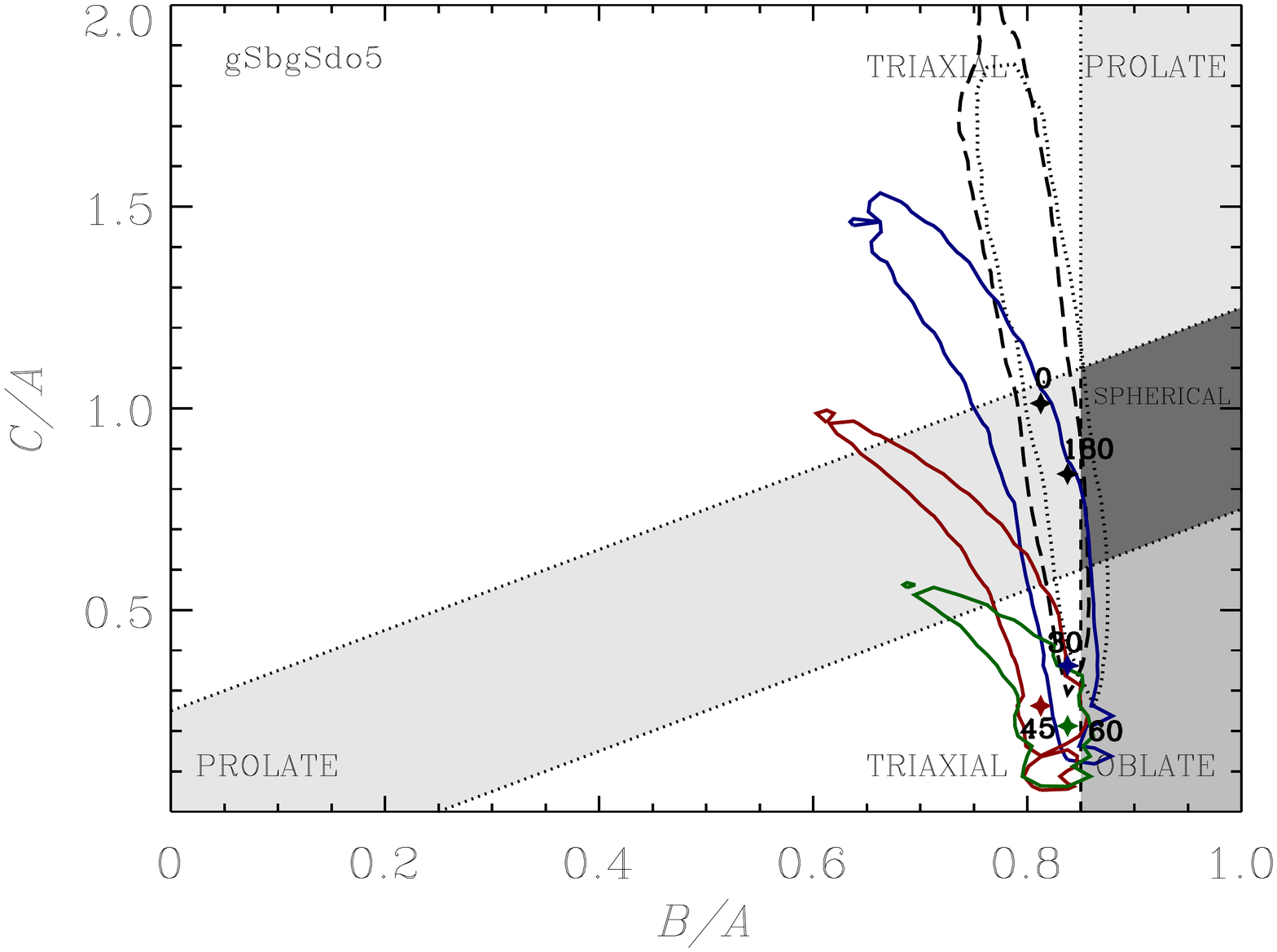}}
\caption{continued.}
\end{figure}

\newpage

\onecolumn

\section{\label{appendix:C}}

\begin{table*}[h!]
\caption{Structural parameters of our CALIFA unbarred galaxies.}
\centering
\begin{adjustbox}{width=0.77\textwidth}
\begin{tabular}{cccccccc}
\hline
\hline
Galaxy 	& $\log(M_{\rm gal}$)		&	$q_{\rm b}$ & $PA_{\rm b}$ &	$q_{\rm d}$ & $PA_{\rm d}$ &	$B/A$ & $C/A$ \\
(1) & (2) & (3) & (4) & (5) & (6) & (7) & (8)\\
\hline
NGC~0001          &   	   10.6   &   0.79  $\pm$    0.02 & 127 $\pm$    3 &   0.64 $\pm$    0.01 &  94 $\pm$    1  & 	 0.71  &  	   0.64  \\
NGC~0160          &         10.9   &   0.71  $\pm$    0.02 &  50 $\pm$    4 &   0.508 $\pm$    0.003 &  48.4 $\pm$    0.2  &      0.96  &       0.61  \\
NGC~0237          &         10.2   &   0.45  $\pm$    0.02 &  47 $\pm$    3 &   0.60 $\pm$    0.01 & 178 $\pm$    1  &      0.29  &       0.41  \\
NGC~0234          &         10.6   &   0.92  $\pm$    0.02 &  77 $\pm$    4 &   0.861 $\pm$    0.003 &  76.8 $\pm$    0.2  &       1.00  &       0.71  \\
NGC~0257          &         10.8   &   0.68  $\pm$    0.01 &  96 $\pm$    2 &   0.621 $\pm$    0.008 &  94.1 $\pm$    0.6  &      0.96  &       0.44  \\
NGC~0496          &         10.3   &   0.86  $\pm$    0.05 &  46 $\pm$    5 &   0.57 $\pm$    0.01 &  33.4 $\pm$    0.8  &      0.91  &       0.81  \\
NGC~0677          &         10.9   &   0.92  $\pm$    0.01 &  31 $\pm$    2 &   0.820 $\pm$    0.008 & 171.2 $\pm$    0.6  &      0.91  &       0.91  \\
NGC~0873          &         10.2   &   0.63  $\pm$    0.01 & 129 $\pm$    2 &   0.840 $\pm$    0.008 & 140.2 $\pm$    0.6  &      0.66  &       0.54  \\
NGC~1070          &         10.8   &   0.97  $\pm$    0.02 &  28 $\pm$    4 &   0.814 $\pm$    0.003 &   1.5 $\pm$    0.2  &      0.96  &       0.94  \\
NGC~1094          &         10.6   &   0.76  $\pm$    0.03 &  97 $\pm$    4 &   0.688 $\pm$    0.008 &  93.1 $\pm$    0.4  &      0.96  &       0.49  \\
NGC~1349          &         10.8   &   0.95  $\pm$    0.02 &  98 $\pm$    3 &   0.88 $\pm$    0.01 &  98 $\pm$    1  &       1.00  &       0.76  \\
NGC~1665          &         10.5   &   0.80  $\pm$    0.01 &  54 $\pm$    2 &   0.559 $\pm$    0.008 &  48.1 $\pm$    0.6  &      0.94  &       0.69  \\
NGC~2476          &         10.5   &   0.71  $\pm$    0.02 & 144 $\pm$    4 &   0.664 $\pm$    0.003 & 148.7 $\pm$    0.2  &      0.96  &       0.46  \\
IC~2341           &         10.8   &   0.55  $\pm$    0.03 &   5 $\pm$    4 &   0.533 $\pm$    0.008 &   1.8 $\pm$    0.4  &      0.94  &       0.29  \\
NGC~2592          &         10.3   &   0.80  $\pm$    0.01 &  58 $\pm$    2 &   0.803 $\pm$    0.008 &  57.9 $\pm$    0.6  &       1.0  &       0.31  \\
NGC~2916          &         10.5   &   0.82  $\pm$    0.02 &   7 $\pm$    4 &   0.651 $\pm$    0.003 &  15.0 $\pm$    0.2  &      0.94  &       0.66  \\
NGC~3106          &         11.0   &   0.96  $\pm$    0.01 & 144 $\pm$    2 &   0.901 $\pm$    0.008 & 135.3 $\pm$    0.6  &       1.00  &       0.81  \\
NGC~3158          &         11.6   &   0.78  $\pm$    0.01 &  70 $\pm$    2 &   0.868 $\pm$    0.008 &  71.0 $\pm$    0.6  &      0.89  &       0.39  \\
UGC~05520         &         	9.5    &   0.34  $\pm$    0.03 & 112 $\pm$    4 &   0.53 $\pm$    0.02 & 100 $\pm$    1  &      0.46  &       0.24  \\
UGC~07012         &         	9.1    &   0.43  $\pm$    0.05 & 159 $\pm$    5 &   0.58 $\pm$    0.01 &  13.8 $\pm$    0.8  &      0.34  &       0.29  \\
IC~0776           &         	9.3    &   0.64  $\pm$    0.05 &  40 $\pm$    5 &   0.54 $\pm$    0.01 &  91.2 $\pm$    0.8  &      0.46  &       0.49  \\
NGC~4711          &         10.3   &   0.66  $\pm$    0.03 &  48 $\pm$    4 &   0.476 $\pm$    0.008 &  41.9 $\pm$    0.4  &      0.84  &       0.54  \\
NGC~5376          &          ...	      &   0.72  $\pm$    0.02 &  59 $\pm$    4 &   0.577 $\pm$    0.003 &  65.3 $\pm$    0.2  &      0.91  &       0.56  \\
UGC~09110         &         10.1   &   0.65  $\pm$    0.02 &  19 $\pm$    3 &   0.44 $\pm$    0.01 &  21 $\pm$    1  &      0.94  &       0.54  \\
NGC~5732          &         	9.9    &   0.72  $\pm$    0.05 &  39 $\pm$    5 &   0.57 $\pm$    0.01 &  40.1 $\pm$    0.8  &       1.00  &       0.59  \\
NGC~5772          &         10.8   &   0.80  $\pm$    0.01 &  38 $\pm$    2 &   0.531 $\pm$    0.008 &  36.9 $\pm$    0.6  &       1.00  &       0.74  \\
NGC~6060          &         10.8   &   0.51  $\pm$    0.01 & 100 $\pm$    2 &   0.435 $\pm$    0.008 & 100.4 $\pm$    0.6  &       1.00  &       0.36  \\
NGC~6155          &         10.1   &   0.45  $\pm$    0.03 & 118 $\pm$    4 &   0.707 $\pm$    0.008 & 146.6 $\pm$    0.4  &      0.41  &       0.39  \\
NGC~6301          &         10.8   &   0.63  $\pm$    0.03 & 110 $\pm$    4 &   0.603 $\pm$    0.008 & 109.6 $\pm$    0.4  &       1.00  &       0.36  \\
NGC~6314          &         11.1   &   0.51  $\pm$    0.02 & 173 $\pm$    3 &   0.56 $\pm$    0.01 & 175 $\pm$    1  &      0.89  &       0.19  \\
NGC~7047          &         10.7   &   0.52  $\pm$    0.03 & 111 $\pm$    4 &   0.491 $\pm$    0.008 & 107.0 $\pm$    0.4  &      0.89  &       0.29  \\
UGC~12224         &         	9.9    &   0.56  $\pm$    0.02 & 103 $\pm$    3 &   0.84 $\pm$    0.01 &  34 $\pm$    1  &      0.51  &        1.04  \\
IC~5309           &         10.2   &   0.41  $\pm$    0.03 &  14 $\pm$    4 &   0.50 $\pm$    0.02 &  26 $\pm$    1  &      0.64  &       0.24  \\
NGC~7653          &         10.5   &   0.89  $\pm$    0.02 &  20 $\pm$    3 &   0.84 $\pm$    0.01 & 164 $\pm$    1  &      0.89  &       0.81  \\
NGC~7782          &         11.1   &   0.71  $\pm$    0.01 & 179 $\pm$    2 &   0.556 $\pm$    0.008 & 176.5 $\pm$    0.6  &      0.96  &       0.56  \\
NGC~5481          &         10.3   &   0.93  $\pm$    0.02 & 114 $\pm$    4 &   0.738 $\pm$    0.003 & 114.8 $\pm$    0.1  &       1.00  &       0.86  \\
UGC~09708         &         10.1   &   0.81  $\pm$    0.05 & 138 $\pm$    6 &   0.76 $\pm$    0.04 & 151 $\pm$    3  &      0.91  &       0.59  \\
UGC~01370         &         10.6   &   0.55  $\pm$    0.05 & 156 $\pm$    5 &   0.43 $\pm$    0.01 & 156.5 $\pm$    0.8  &      0.94  &       0.41  \\
NGC~5145          &         	9.9    &   0.58  $\pm$    0.01 &  88 $\pm$    2 &   0.807 $\pm$    0.008 &  56.0 $\pm$    0.6  &      0.54  &       0.56  \\
MCG~$-$01$-$52$-$012    &         10.3   &   0.47  $\pm$    0.02 &  86 $\pm$    3 &   0.80 $\pm$    0.01 &  43 $\pm$    1  &      0.39  &       0.54  \\
UGC~09837         &         	9.1    &   0.61  $\pm$    0.05 &  31 $\pm$    5 &   0.81 $\pm$    0.01 & 137.7 $\pm$    0.8  &      0.59  &        1.26  \\
NGC~2526          &         10.2   &   0.68  $\pm$    0.05 & 142 $\pm$    5 &   0.51 $\pm$    0.01 & 130.9 $\pm$    0.8  &      0.76  &       0.51  \\
MCG~$+$09$-$22$-$053    &         	9.4    &   0.79  $\pm$    0.05 &  93 $\pm$    5 &   0.77 $\pm$    0.01 & 127.9 $\pm$    0.8  &      0.79  &       0.64	\\
\hline
\end{tabular}
\end{adjustbox}
\tablefoot{(1) Galaxy name. (2) Stellar mass of the galaxy from \citet{Walcher2014}. 
(3), (4) Apparent axial ratio and position angle of the bulge.
(5), (6) Apparent axial ratio and position angle of the disk.
(7), (8) Most probable intrinsic axial ratios $B/A$ and $C/A$ of the bulge.}
\label{tab:appendix:C1}
\end{table*}

\begin{table*}[h!]
\caption{As in Table~\ref{tab:appendix:C1}, but for barred galaxies.}
\centering
\begin{adjustbox}{width=0.77\textwidth}
\begin{tabular}{cccccccc}
\hline
\hline
Galaxy 	& $\log(M_{\rm gal}$)		&	$q_{\rm b}$ & $PA_{\rm b}$ &	$q_{\rm d}$ & $PA_{\rm d}$ &	$B/A$ & $C/A$ \\
(1) & (2) & (3) & (4) & (5) & (6) & (7) & (8)\\
\hline
NGC~0171      &   	   10.4      &   0.73 $\pm$   0.02 & 124 $\pm$   3 &   0.885 $\pm$   0.004 & 	  97.6	$\pm$   0.3	&  		0.71   &     0.76   \\
NGC~0309      &         10.7      &   0.88 $\pm$   0.03 & 136 $\pm$   3 &   0.894 $\pm$   0.007 & 	 108.4	$\pm$   0.3 &       0.91   &        0.61   \\
NGC~0364      &         10.6      &   0.88 $\pm$   0.05 &  45 $\pm$   6 &   0.73  $\pm$   0.01 &  	 33.3 	$\pm$   0.7 &       0.93   &        0.76   \\
NGC~0551      &         10.6      &   0.64 $\pm$   0.05 & 126 $\pm$   6 &   0.44  $\pm$   0.01 &  	135.9 	$\pm$   0.7 &       0.78   &        0.53   \\
NGC~0842      &         10.8      &   0.64 $\pm$   0.02 & 137 $\pm$   3 &   0.525 $\pm$   0.004 & 	 145.1	$\pm$   0.3 &       0.86   &        0.46   \\
NGC~1666      &         10.5      &   0.88 $\pm$   0.02 & 138 $\pm$   3 &   0.880 $\pm$   0.004 & 	 147.5 	$\pm$   0.3 &       0.96   &        0.41   \\
NGC~1667      &         10.7      &   0.57 $\pm$   0.03 & 171 $\pm$   3 &   0.687 $\pm$   0.007 & 	 172.1 	$\pm$   0.3 &       0.81   &        0.26   \\
UGC~03253     &         10.4      &   0.65 $\pm$   0.05 &  92 $\pm$   6 &   0.60 $\pm$   0.01 &  	 78.4 	$\pm$   0.7 &       0.78   &        0.43   \\
NGC~2486      &         10.6      &   0.83 $\pm$   0.04 &  85 $\pm$   5 &   0.591 $\pm$   0.009 & 	  90.7	$\pm$   0.5 &       0.93   &        0.76   \\
UGC~04145     &         10.6      &   0.60 $\pm$   0.05 & 135 $\pm$   6 &   0.50 $\pm$   0.01 &  	138.2 	$\pm$   0.7 &       0.93   &        0.48   \\
NGC~2572      &         10.9      &   0.59 $\pm$   0.07 & 126 $\pm$  13 &   0.43 $\pm$   0.02 &  	137.4 	$\pm$   0.9 &       0.51   &        0.33   \\
NGC~2880      &         10.4      &   0.79 $\pm$   0.02 & 129 $\pm$   3 &   0.571 $\pm$   0.003 & 	 143.2	$\pm$   0.1 &       0.86   &        0.66   \\
NGC~3381      &          9.6      &   0.70 $\pm$   0.05 &  80 $\pm$   6 &   0.83 $\pm$   0.01 &  	 45.2 	$\pm$   0.7 &       0.76   &        0.28   \\
NGC~4185      &         10.6      &   0.67 $\pm$   0.02 & 173 $\pm$   3 &   0.666 $\pm$   0.004 & 	 167.0 	$\pm$   0.3 &       0.91   &        0.31   \\
NGC~4210      &         10.3      &   0.75 $\pm$   0.02 &  78 $\pm$   3 &   0.731 $\pm$   0.004 & 	  94.1 	$\pm$   0.3 &       0.83   &        0.38   \\
NGC~4961      &          9.6      &   0.67 $\pm$   0.04 & 111 $\pm$   5 &   0.692 $\pm$   0.009 & 	  99.9 	$\pm$   0.5 &       0.83   &        0.31   \\
NGC~5056      &         10.6      &   0.58 $\pm$   0.05 &  97 $\pm$   6 &   0.55 $\pm$   0.01 &  	179.4 	$\pm$   0.7 &       0.46   &        0.96   \\
NGC~5157      &         11.1      &   0.73 $\pm$   0.05 & 114 $\pm$   6 &   0.78 $\pm$   0.01 &  	105.7 	$\pm$   0.7 &       0.91   &        0.28   \\
NGC~5473      &         10.6      &   0.92 $\pm$   0.01 & 137 $\pm$   2 &   0.787 $\pm$   0.003 & 	 155.0 	$\pm$   0.1 &       0.93   &        0.81   \\
IC~0994       &         11.1      &   0.78 $\pm$   0.05 &  19 $\pm$   6 &   0.51 $\pm$   0.01 &  	 14.6 	$\pm$   0.7 &       0.91   &        0.71   \\
NGC~5602      &         10.5      &   0.83 $\pm$   0.08 & 163 $\pm$   8 &   0.52 $\pm$   0.03 &  	167 	$\pm$   2   &        1.00   &       0.81    \\
NGC~5720      &         10.8      &   0.82 $\pm$   0.05 & 125 $\pm$   6 &   0.65 $\pm$   0.01 &  	129.2 	$\pm$   0.7 &        1.00   &       0.73    \\
NGC~5735      &         10.1      &   0.78 $\pm$   0.05 &  73 $\pm$   6 &   0.90 $\pm$   0.01 &  	 32.6 	$\pm$   0.7 &       0.78   &        0.36   \\
UGC~09492     &         11.1      &   0.65 $\pm$   0.05 &  47 $\pm$   6 &   0.45 $\pm$   0.01 &  	 54.2 	$\pm$   0.7 &       0.78   &        0.48   \\
IC~4534       &         10.7      &   0.65 $\pm$   0.04 & 158 $\pm$   5 &   0.564 $\pm$   0.009 & 	 162.9 	$\pm$   0.5 &       0.93   &        0.48   \\
NGC~5888      &         11.2      &   0.69 $\pm$   0.04 & 154 $\pm$   5 &   0.596 $\pm$   0.009 & 	 153.2 	$\pm$   0.5 &       0.96   &        0.51   \\
UGC~09777     &         10.2      &   0.95 $\pm$   0.05 & 129 $\pm$   5 &   0.61 $\pm$   0.02 &  	145.9 	$\pm$   0.9 &        1.00   &       0.93    \\
NGC~6278      &         10.7      &   0.81 $\pm$   0.02 & 123 $\pm$   3 &   0.531 $\pm$   0.004 & 	 126.8 	$\pm$   0.3 &       0.96   &        0.73   \\
NGC~6941      &         10.9      &   0.68 $\pm$   0.05 & 118 $\pm$   6 &   0.72 $\pm$   0.01 &  	130.5 	$\pm$   0.7 &       0.86   &        0.23   \\
UGC~11649     &         10.4      &   0.82 $\pm$   0.05 & 132 $\pm$   6 &   0.86 $\pm$   0.01 &  	 71.8 	$\pm$   0.7 &       0.81   &         1.00   \\
NGC~7321      &         10.9      &   0.63 $\pm$   0.05 &  15 $\pm$   6 &   0.65 $\pm$   0.01 &  	 22.2 	$\pm$   0.7 &       0.86   &        0.33   \\
UGC~12185     &         10.5      &   0.71 $\pm$   0.05 & 141 $\pm$   5 &   0.46 $\pm$   0.02 &  	155.0 	$\pm$   0.9 &       0.73   &        0.58   \\
NGC~7591      &         10.7      &   0.85 $\pm$   0.05 & 167 $\pm$   6 &   0.48 $\pm$   0.01 &  	148.4 	$\pm$   0.7 &       0.83   &        0.76   \\
NGC~7671      &         10.7      &   0.83 $\pm$   0.02 & 142 $\pm$   3 &   0.596 $\pm$   0.004 & 	 135.1 	$\pm$   0.3 &       0.93   &        0.71   \\
NGC~7716      &         10.3      &   0.51 $\pm$   0.04 &  56 $\pm$   5 &   0.822 $\pm$   0.008 & 	  39.4 	$\pm$   0.4 &       0.51   &        0.48   \\
UGC~04455     &         10.9      &   0.73 $\pm$   0.08 & 174 $\pm$   8 &   0.74 $\pm$   0.03 &  	 11 	$\pm$   2   &       0.81   &        0.41   \\
NGC~6977      &         10.9      &   0.94 $\pm$   0.05 & 171 $\pm$   6 &   0.83 $\pm$   0.01 &  	152.7 	$\pm$   0.7 &       0.96   &        0.88   \\
UGC~12250     &         11.1      &   0.78 $\pm$   0.04 &  13 $\pm$   5 &   0.626 $\pm$   0.009 & 	  12.7 	$\pm$   0.5 &        1.00   &       0.63    \\
NGC~5947      &         10.6      &   0.88 $\pm$   0.04 &  39 $\pm$   5 &   0.811 $\pm$   0.009 & 	  63.7 	$\pm$   0.5 &       0.88   &        0.71   \\
NGC~2767      &         10.8      &   1.00 $\pm$   0.04 & 178 $\pm$   5 &   0.733 $\pm$   0.009 & 	 169.6 	$\pm$   0.5	&        1.0   &        0.98   \\
\hline
\end{tabular}
\end{adjustbox}
\tablefoot{(1) Galaxy name. (2) Stellar mass of the galaxy from \citet{Walcher2014}. 
(3), (4) Apparent axial ratio and position angle of the bulge.
(5), (6) Apparent axial ratio and position angle of the disk.
(7), (8) Most probable intrinsic axial ratios $B/A$ and $C/A$ of the bulge.}
\label{tab:appendix:C2}
\end{table*}

\end{appendix}

\end{document}